\documentclass[aps,notitlepage,twocolumn,nofootinbib,pra,10pt]{revtex4-2}
\usepackage{amsmath,amsfonts,amssymb,amsthm,bbm,graphicx,enumerate,times,mathtools,hyperref,physics,marvosym,wasysym,stmaryrd,wrapfig}
\hypersetup{colorlinks,linkcolor=blue,citecolor=violet,urlcolor=blue}
\usepackage{subcaption} % For subfigures
\usepackage{float}
\usepackage[ruled,longend]{algorithm2e}
\usepackage{fancyref}
\usepackage{etoolbox}
\usepackage{appendix,soul}
\usepackage[dvipsnames]{xcolor}
\usepackage[noabbrev,capitalise]{cleveref}
\usepackage{array}     % For improved column alignment
\usepackage{multirow}  % For spanning multiple rows or columns
\usepackage{siunitx}

\creflabelformat{equation}{#2#1#3}

\begin{document}
%--------------------------------------------%
\title{Near-Term Spin-Qubit Architecture Design via Multipartite Maximally-Entangled States}
\author{N.~Paraskevopoulos$^{1,2}$*}
\author{M.~Steinberg$^{1,2}$*}
\author{B.~Undseth$^{1,3}$}
\author{A.~Sarkar$^{1,2}$}
\author{L. M. K.~Vandersypen$^{1,3}$}
\author{X.~Xue$^{1,3}$}
\author{S.~Feld$^{1,2}$}
\affiliation{$^{1}$QuTech, Delft University of Technology, Delft, The Netherlands}
\affiliation{$^{2}$Quantum and Computer Engineering Department, Delft University of Technology, Delft, The Netherlands}
\affiliation{$^{3}$Kavli Institute of Nanoscience, Delft University of Technology, Delft, The Netherlands}

\date{\today}

%--------------------------------------------%
\begin{abstract}
The design and benchmarking of quantum computer architectures traditionally rely on practical hardware restrictions, such as gate fidelities, control, and cooling. At the theoretical and software levels, numerous approaches have been proposed for benchmarking quantum devices, ranging from, inter alia, quantum volume to randomized benchmarking. In this work, we utilize the quantum information-theoretic properties of multipartite maximally-entangled quantum states, in addition to their correspondence with quantum error correction codes, permitting us to quantify the entanglement generated on near-term bilinear spin-qubit architectures. For this aim, we introduce four metrics which ascertain the quality of genuine multipartite quantum entanglement, along with circuit-level fidelity measures. As part of the task of executing a quantum circuit on a device, we devise simulations which combine expected hardware characteristics of spin-qubit devices with appropriate compilation techniques; we then analyze three different architectural choices of varying lattice sizes for bilinear arrays, under three increasingly realistic noise models. We find that if the use of a compiler is assumed, sparsely-connected spin-qubit lattices can approach comparable values of our metrics to those of the most highly-connected device architecture. Even more surprisingly, by incorporating crosstalk into our last noise model, we find that, as error rates for crosstalk approach realistic values, the benefits of utilizing a bilinear array with advanced connectivity vanish. Our results highlight the limitations of adding local connectivity to near-term spin-qubit devices, and can be readily adapted to other qubit technologies. The framework developed here can be used for analyzing quantum entanglement on a device before fabrication, informing experimentalists on concomitant realistic expectations. 
\end{abstract}

\maketitle

\def\thefootnote{*}\footnotetext{These authors contributed equally to this work. The corresponding author can be reached at \url{m.a.steinberg@tudelft.nl}.}\def\thefootnote{\arabic{footnote}}

% \tableofcontents

%--------------------------------------------%
\section{Introduction}

Quantum computers promise to solve certain classes of problems more efficiently than classical computers \cite{nielsenchuang}. While great progress has been made in effectively scaling the physical size of a quantum system, it is not yet clear how to optimally arrange and connect qubits on a lattice, given a particular qubit technology. Indeed, connectivity on solid-state quantum hardware is often determined by the ease of the fabrication process. New manufacturing techniques and qubit technologies are constantly being explored, and the design and benchmarking of a quantum device traditionally takes into account hardware-specific variables such as control electronics, power dissipation, calibration, and crosstalk suppression. However, as a quantum device will ideally be fabricated for the execution of real quantum algorithms involving the generation of highly-entangled many-body quantum states, benchmarking and design methods should also factor in how well a quantum device can generate \emph{genuine multipartite entanglement} (i.e., entanglement characterized by quantum correlations between each party in a multiparty state \cite{mme_states_review,guhne2009entanglement}).

How such entanglement is generated varies substantially between different flavors of quantum hardware. As a concrete example, superconducting qubits with nearest-neighbor interactions distribute entanglement very differently from trapped-ions with all-to-all connectivity. Furthermore, different architectures of a specific quantum hardware platform may be better or worse at distributing entanglement; neutral atoms \cite{bluvstein2022quantum}, trapped-ions \cite{quantinuum_racetrack} and semiconductor spins \cite{de2024high,schreiber2018toward} can be physically shuttled for dynamic qubit connectivity, enabling more flexibility in certain respects. The ability for these platforms to generate multipartite entanglement therefore depends greatly on the precise architecture, the quality of the primitive operations, and the compilation of the quantum circuit to be executed.

Several methods for benchmarking quantum computer performance currently exist; these range from \emph{randomized benchmarking} \cite{randomized2,randomized3} and various forms of \emph{quantum tomography} \cite{process_tomography}, to other methods such as calculating the \emph{estimated success probability} (ESP) \cite{SpinQ,ArtA,schmid2024computational,nishio2020extracting} used at the compiler level, \emph{quantum volume} \cite{cross2019validating} and its derivatives \cite{wack2021quality,blume2020volumetric}, as well as through demonstrations of \emph{quantum advantage} \cite{arute2019quantum} and application-level benchmarks \cite{sebastian1}. Although there is no shortage of benchmarking techniques, neither is there a current consensus regarding the performance qualification of a quantum device, especially when taking into account different types of qubit technologies, architectural schemes, as well as available elementary gate sets and the topological-graph properties of quantum algorithms \cite{medina_qmi_paper,steinberg2022topological}. In this way, all current benchmarking procedures and protocols exhibit their own advantages and disadvantages; as an example, \emph{gate set tomography} is known to provide self-consistent characterizations of physical gates; unfortunately, most techniques are known to scale exponentially in the number of qubits \cite{nielsen2021gate}. Another example, quantum volume, takes into account many essential metrics that factor into processor performance: circuit depth, gate fidelity, qubit count, and qubit connectivity; however, the circuits crafted using this technique are usually not representative of realistic quantum algorithms \cite{cross2019validating}. This fact has spurred interest in the development of alternative methods for benchmarking at the algorithmic level for existing quantum processors \cite{ketgpt, tomesh2022supermarq, quetschlich2023mqt, proctor2024benchmarking, liu2023user, lubinski2023application}.  

It was first proposed in \cite{quantumcirc_mme_states} that one might be able to utilize \emph{multipartite maximally-entangled} (MME) states, in particular, \emph{absolutely maximally-entangled} (AME) quantum states, as a suitable and generalizable benchmark for the quality of entanglement generated on a quantum device \cite{quantumcirc_mme_states}. The main reason for utilizing AME states lies in the implementation details: the authors describe the preparation of highly-entangled states as difficult tasks for today's devices, mainly due to the genuine multipartite nature of the quantum entanglement required among all subsystems \cite{mme_states_review}. Additionally, utilizing AME states carries several other advantages. Firstly, it is known that AME states and their less highly-entangled counterparts, \emph{k-uniform} states, exhibit an exact mathematical duality to \emph{maximal distance separable} (MDS) codes; MDS codes are known to exhibit the highest distance allowed by the \emph{quantum Singleton bound}, and are therefore optimal, in terms of their code-theoretic properties \cite{raissi1,raissi2,mds_codes_ame}. Benchmarking or designing a quantum device with respect to these many-body states is then directly related to understanding how resilient logical qubits can be fashioned on a device. Secondly, AME states are, by definition, the most highly-entangled quantum states possible for a given number of qudits. In this way, using AME and other highly-entangled MME states for entanglement characterization on a noisy device provides a benchmark that employs the entire multipartite Hilbert space available in order to generate a state, rather than a subset of qubits. This relationship was illustrated in a recent experiment measuring objects related to \emph{quantum weight enumerators} of AME and k-uniform states, and conveyed a direct method for entanglement characterization via a relationship to the \emph{quantum concurrence}, a known entanglement monotone \cite{daniel_miller}. 

In the practical implementation of a quantum algorithm, one must also consider not just the difficulty of realizing the quantum circuit generating a particular output state, but just as importantly, the specific compilation requirements arising from the choice of qubit technology, architectural connectivity, and the constraints that follow therein. The challenge of efficient quantum compilation is essential to assess, even at the level of \emph{logical qubits}, since it is understood that concepts from \emph{quantum error correction} (QEC) currently offer the only feasible way by which a quantum computer could be scaled to the large system sizes needed in order to execute useful quantum algorithms \cite{lidar2013quantum}. Taking stock of these points, it seems reasonable to incorporate MME- and QEC-motivated design benchmarks which factor in the benefits (as well as the burden) of quantum compilation, unifying multiple angles of entanglement characterization in a framework. 

In this work, we propose a framework consisting of four quantum information-theoretic and compilation-motivated measures for designing and benchmarking quantum devices; our purpose here is not to propose a complete parameterization of device design, but rather to introduce mathematical concepts that can help to inform the design process. The central theme of our choice of measures lies in the fact that they exploit the highly-entangled nature of MME states and their relationship to small quantum error correction codes. Although the methods we introduce can be adapted to any qubit technology, we assess example designs for near-term \emph{spin-qubit} architectures, focusing on those with \emph{bilinear} qubit layout, but with varying connectivities, lattice array sizes, and ultimately, the qubit density (as it relates to the number of empty quantum dots on the device), following recent experimental trends \cite{hsiao2024exciton, crawford2023compilation, siegel2024towards, xue2022performance}. As the spin-qubit community is on the verge of performing its first fault-tolerance and QEC demonstrations beyond the repetition code \cite{gozde_article_floquet_qec,morello_logical,siegel2024towards,dawn_bartlett,phase_flip_spin,hetenyi2024tailoring,takeda2022quantum_phase_2,silicon_surface_code_architecture,silicon_architecture}, it is timely and prudent to consider just how much of a role local connectivity could play in the realization of near-term QEC experiments for spin qubit technology, especially if we compare the relative ease of performing quantum compilation techniques with the difficulty of fabricating entirely new processors with advanced connectivities. 

More explicitly, we consider four main simulations. Firstly, we consider the generation of an AME state of six qubits, which is similar to what was carried out in \cite{quantumcirc_mme_states}, and was first discovered in \cite{borras2007multiqubit}; we take the additional step of calculating a quantity known as the \emph{Bell operator}, which is widely understood as a metric for gauging the \emph{quantumness} of qubit-qubit correlations among the parties of a genuinely entangled multipartite state. Our second test involves the usage of an MME state related to AME states: the \emph{k-uniform} state. Here, we map the state to the smallest error-detecting surface code (i.e., the $\llbracket 4,1,2\rrbracket$ surface code with three ancilla qubits, as described in \cite{experiment2,experiment3}), and evaluate the logical qubit's failure rate over many cycles of stabilizer measurements; as it is known that AME states of four qubits do not exist \cite{ame_existence4}, the k-uniform state that we have chosen also maximizes the entanglement possible in a four-party Hilbert space. Thirdly, we invoke a modified version of the \emph{estimated success probability} (ESP), a standard quantum compilation measure used to grade the worst-case scenario of circuit performance on a device; our modification takes into account the \emph{decoherence-induced} noise that is expected on near-term spin-qubit devices. Finally, the fourth test employs an \emph{entropic divergence measure} known as the \emph{tripartite mutual information}; this is known to indicate the degree to which quantum entanglement is distributed globally versus locally in a many-body quantum system undergoing unitary evolution and also periodic measurements \cite{tripartite1,tripartite2}. 

Our main contributions are as follows. Firstly, we develop a framework for critiquing the design of spin-qubit architectures, utilizing the previously-mentioned quantum-information-based objects, together with state-of-the-art spin-qubit compilation techniques \cite{SpinQ,beSnake}. Our goal is to evaluate several architectural proposals for bilinear spin-qubit arrays of varying lattice size; we fill these arrays with seven qubits, and then decrease the qubit density of the arrays by increasing the number of quantum dots in the structure. In the semiconductor spin-qubit field, the entire lattice is usually filled with qubits; however, our study shows large benefits to lowering the qubit density, allowing for increased compilation flexibility via \emph{qubit-shuttling} techniques \cite{yoneda2021coherent,de2024high,van2024coherent,struck2024spin}. The innate flexibility in spin-qubit devices lies in stark contrast to connectivity-constrained devices, such as superconducting qubit technology, for which SWAP gates are needed. 

Secondly, we perform a detailed analysis of the aforementioned near-term devices, providing guidance on which of the architectures considered is best suited for small QEC and \emph{fault-tolerance} experiments with bilinear spin-qubit arrays. Our results show that, if we assume the inclusion of hardware-specific quantum compilation techniques, then it is possible for sparsely-connected devices to achieve values in our metric that are comparable to those of the most highly-connected spin-qubit device in our study. Even further, our simulations with crosstalk indicate that the benefits of adding complex local connectivity are outweighed by increasing crosstalk effects, in essence providing a limitation for architectural connectivity in spin-qubit devices. Our work strongly implies that there are constraints on how much local connectivity can benefit a spin-qubit device at the moment of generating genuine multipartite entanglement, and that appropriate compilation techniques can compensate for the lack of connectivity in an architecture. Our results reveal that efficient quantum compilation techniques can aid in the realization of small quantum error correction experiments better than advanced connectivity in the actual spin-qubit device. 

This article is organized as follows. \cref{section:background} introduces multipartite maximally-entangled states, their relationship to quantum error correction codes, and the entanglement metrics that we propose in our framework (\cref{section:mme_qec,section:k_uni_surface_code,section:monitored_quantum_circs}), as well as our modified version of estimated success probability that we employ in our study (\cref{section:esp_background}). We additionally cover the basics of quantum compilation (\cref{section:compilation_background}) and some of the details regarding spin-qubit architectures (\cref{section:spin_qubit_archs}). \cref{section:results} showcases the concrete results obtained for each of the simulation experiments; we follow up by analyzing and discussing each of these in \cref{section:discussion}. In \cref{section:conclusion}, we provide final comments and summarize potential future directions.

%--------------------------------------------%
\section{Background} \label{section:background}

In what follows, we introduce the four main metrics utilized in this work. Additionally, we provide an introductory section on \emph{quantum compilation} and its role in our investigation, as well as a review of the the basics pertaining to spin-qubit architectures.

%--------------------------------------------%
\subsection{Multipartite Maximally-Entangled States \& Quantum Error Correction} \label{section:mme_qec}

\emph{Multipartite maximally-entangled} (MME) states are generalized, highly-entangled many-body quantum systems, for which it is understood that certain reductions of these states become \emph{maximally mixed} \cite{mme_states_review}. Such quantum systems have been shown to exhibit intimate connections with many aspects of quantum information theory such as quantum error correction, as well as quantum communication \cite{theory_secret_share,quantum_secret_share,gottesman1997_thesis} and toy models of quantum gravity theories such as the AdS/CFT correspondence \cite{happy_codes,harris_css_block_perfect,steinberg_htn_codes,farrelly_tn_codes,steinberg2024far}. Although formulating a completely exhaustive and general framework for all MME states is very complex, much progress has been made by studying subsets of states. Some examples of these subsets are: \emph{graph states} \cite{eisert2}; \emph{tensor network states} \cite{orus}; and certain classes of quantum error correction codes, e.g. \emph{maximal distance separable} (MDS) codes \cite{raissi1,raissi2,mds_codes_ame}.

In the simplest case, the Bell state $\ket{\Psi^{+}_{2}} = \frac{1}{\sqrt{2}}(\ket{00}+\ket{11})$ is an example of a maximally-entangled quantum state. Performing a partial trace on either of the two subsystems (which we label $\mathrm{A}$) results in a maximally-mixed quantum state of the form 

\begin{equation}
\text{Tr}_{\mathrm{A}}\big[ \ket{\Psi^{+}_{2}}\bra{\Psi^{+}_{2}}\big] \propto \mathbb{I}_{2}~. \label{eq:k_uni_condition} 
\end{equation}

One can also define generalized, multipartite quantum systems, in which the potential bipartitions of such a state and the associated reduced density matrix yield a maximally mixed entanglement spectrum \cite{mme_states_review}. 

A prominent example of a generalized multipartite maximally entangled state is the \emph{absolutely maximally-entangled} (AME) state, which is an $n$-qudit state $\ket{\psi}$, defined in $\mathcal{H}^{n}_{q} := \mathbb{C}^{\otimes n}_{q}$ (wherein $q$ denotes local dimension), for which the following condition holds:

\begin{equation}
\rho_{s} = \text{Tr}_{s^{c}} \big[ \ket{\psi}\bra{\psi} \big] \propto \mathbb{I}~, \label{eq:ame}
\end{equation}

where the subset $s \subset \{1 \cdots n\}$, $|s| = \left \lfloor{\frac{n}{2}}\right \rfloor$, and $s^{c}$ denotes the complementary set of subsystems to $s$. For the present work, we use the shorthand form AME($n,q$) to refer to an $n$-qudit, $q$ local dimension AME state of interest. 

Furthermore, AME states are closely related to quantum error correction codes. In particular, it was shown in \cite{raissi2} that an exact mathematical duality exists between a given AME state, AME($n,q$), and a \emph{maximal distance separable} (MDS) code with parameters $\llbracket n,k,d\rrbracket_{q}$, where $k,d$ represent the number of encoded logical qubits as well as the distance of the code. In particular, it was shown in \cite{raissi1} that \emph{code shortening} techniques could be utilized in order to generate entire families of AME states with provably minimal support.

Constructions of particular AME states were provided in \cite{quantumcirc_mme_states}, and subsequent translations to the circuit level were given. These constructions mainly relied on mapping some AME($n,q$) to a graph state, which are defined as an $n$-partite pure quantum state composed of $n$ vertices $\mathcal{V} = \{v_{1} \cdots v_{i} \cdots v_{n}\}$, and edges $\mathcal{E} = \{e_{ij} = \{v_{i},v_{j}\} \}$. Each graph has an associated \emph{adjacency matrix} $A$, whose entries $A_{ij}$ satisfy $A_{ij} = 1$ if an edge exists; otherwise, the entry's value is zero (self interactions are forbidden in this formalism). As we will only be utilizing the AME(6,2) state in the present work, we define an $n$-qubit graph state as 

\begin{equation}
\ket{G} = \prod^{n}_{i<j}\text{CZ}^{A_{ij}}_{ij}\big[\ket{+}\big]^{\otimes n}~.
\label{eq:qubit_graph_state_ame}
\end{equation}

Note that AME states can be defined using the graph state formalism for any local dimension \cite{helwig2013absolutely,alsina_our_ame62,mds_codes_ame}, with the well-known \emph{cluster states} as a specific case \cite{cluster}.

The AME(6,2) state, given in \cite{quantumcirc_mme_states,borras2007multiqubit}, possesses the following stabilizers:

\begin{align}
XZIIZZ~, \\
ZXZZII~, \\
IZXZIZ~, \\
IZZXZI~, \\
ZIIZXZ~, \\
ZIZIZX~.
\end{align}

The explicit form of the state is 

\begin{align}
\ket{\Omega_{6,2}} = \frac{1}{4}\big[ \ket{000}(\ket{\texttt{+-+}} +\ket{\texttt{-+-}}) - \nonumber \\ 
\ket{001}(\ket{\texttt{+--}}-\ket{\texttt{-++}}) + \nonumber\\
\ket{010}(\ket{\texttt{++-}}-\ket{\texttt{--+}}) - \nonumber\\
\ket{011}(\ket{\texttt{+++}}+\ket{\texttt{---}}) - \nonumber\\
\ket{100}(\ket{\texttt{+++}}-\ket{\texttt{---}}) - \nonumber\\
\ket{101}(\ket{\texttt{++-}}+\ket{\texttt{--+}}) - \nonumber\\
\ket{110}(\ket{\texttt{+--}}+\ket{\texttt{-++}}) - \nonumber\\
\ket{111}(\ket{\texttt{+-+}}-\ket{\texttt{-+-}})\big]~.
\end{align}

\begin{figure}
\centering
\includegraphics[width=0.8\columnwidth]{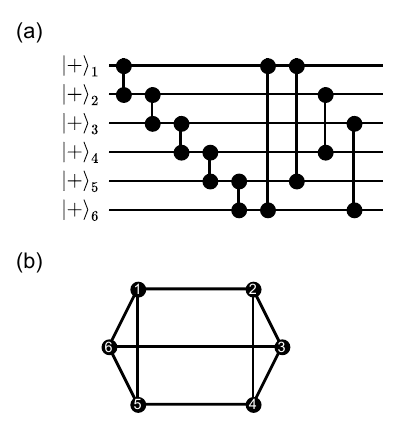}
\caption{Generating circuit for the AME(6,2) state. (a) depicts the circuit-level implementation that was utilized for this work, and (b) is a graphical representation of the resulting \emph{graph state} and its qubit-qubit entanglement correlations. 2-qubit gates are shown as $CZ$ gates in (a).}
\label{fig:ame_generating_circuit}
\end{figure}

The circuit that generates the AME(6,2) is shown in \cref{fig:ame_generating_circuit}. Here, the six data qubits in the AME(6,2) circuit are initialized in the $\ket{+}$ state and undergo the commensurate CZ gates shown in (a), resulting in entanglement correlations among the six parties, as shown in (b). The circuit and graph state form displayed follow from the definition of AME qubit graph states in \cref{eq:qubit_graph_state_ame}.

Using graph states, one can derive the famous \emph{Bell inequalities}, which are known to provide bounds on the correlations allowed by so-called \emph{local-hidden-value} (LHV) theories. Quantum states exhibiting entanglement correlations are known to violate these inequalities, and graph states can be used to directly derive this fact \cite{bell_operator,eisert2}. Central to this derivation lies in the calculation of the \emph{Bell operator} $B(G)$, which is defined as

\begin{equation}
B(G) = \sum_{i}^{2^{n}} S_{i}~,
\end{equation}

where $S_{i}$ is a \emph{stabilizer} of graph state $G$. The stabilizer operator is defined as 

\begin{equation}
S_{i} = X_{i} \bigotimes_{j \in A_{ij}} Z_{j}~,
\end{equation}

which is typical for a graph state. 

One can verify via direct computation that the number of stabilizers for a given graph state is exactly $2^{n}$ \cite{eisert2}. In this way, the Bell operator provides useful information on the \emph{quantumness} of correlations between neighboring qubits; the Bell operator itself is bounded from above by $2^{n}$, which represents purely quantum correlations, and from below by $2^{(n-1)/2}$, signifying purely classical correlations \cite{bell_operator,quantumcirc_mme_states}. For the AME(6,2) state, the average expectation value of the Bell operator is found to be within the bounds $4 \leq \langle B \rangle \leq 64$.

It is well-known that AME states do not exist for all local dimensions, given a number of qudits; as an example, AME qubit states only exist for $n = 2, 3, 5$, and  $6$ parties \cite{ame_existence4,ame_existence7,huber2018bounds_ameexistence}. As a result, less highly-entangled quantum states known as \emph{k-uniform} states that generalize the AME condition were introduced \cite{alsina_our_ame62}; these states in turn can be defined for any number of parties, for any local dimension. k-uniform states are defined as adhering to the same constraint as in \cref{eq:ame}, but, in contrast, the requirement $k = |s| \leq \left \lfloor{\frac{n}{2}}\right \rfloor$ is relaxed. As an example, the famous \emph{Greenberger–Horne–Zeilinger} (GHZ) state constitutes a 1-uniform state, as any reduction to one qubit results in a maximally-mixed state \cite{greenberger1989going,nielsenchuang}. In \cite{raissi3}, a generalized formalism for constructing stabilizer quantum error correction codes from k-uniform states was introduced.

%--------------------------------------------%
\subsection{k-Uniform Quantum States \& the Surface Code} \label{section:k_uni_surface_code}

The \emph{surface code} is a family of $\llbracket L^{2} + (L-1), 1, L\rrbracket$ stabilizer quantum error correction codes (with $L$ signifying the length of a square lattice) that originally arose from the celebrated \emph{toric code} \cite{fowler2012surface,terhal2015quantum,kitaev2003fault}. Originally arranged on a toroidal lattice, it was discovered thereafter that such a lattice orientation was unnecessary, and planar versions were proposed \cite{bravyi1998quantum}, which eventually became known as the \emph{surface code}. These codes have been proposed and experimentally realized \cite{google1,google2,experiment1,experiment2,experiment3} as the current leading candidate for universal, fault-tolerant logical quantum computation \cite{terhal2015quantum,fowler2012surface}, owing to their resilience against many varieties of noise, the development of \emph{lattice-surgery} techniques \cite{lattice_surgery1}, ease of experimental implementation via a planar layout \cite{fowler2012surface}, and \emph{magic-state distillation} protocols \cite{msd0} which provide for universal logical gate sets. Many variants of the surface code have been studied and proposed, including the \emph{rotated} \cite{rotated1} and \emph{XZZX} \cite{XZZX} versions, among others \cite{3Dsurface_codes,XY}. 

The smallest representative of the \emph{rotated surface code} family, with parameters $\llbracket d^{2}, 1, d \rrbracket$, is the $\llbracket 4,1,2 \rrbracket$ \emph{quantum error-detecting} surface code, so named due to the distance of the code. The stabilizers of the code take the form:

\begin{align}
XXXX~, \\
IZIZ~, \\
ZIZI~,
\end{align}

and the logical operators are $IXIX, IIZZ$. Here we use the convention that $XX = X \otimes X$ for brevity. The code states are defined as:

\begin{align}
\ket{\bar{0}} = \frac{1}{\sqrt{2}}\big[ \ket{0000} + \ket{1111}\big]~, \\
\ket{\bar{1}} = \frac{1}{\sqrt{2}}\big[ \ket{0101} + \ket{1010}\big]~.
\end{align}

Using \cref{eq:k_uni_condition}, one can easily check that the four-qubit state described above constitutes an example of a \emph{planar} 2-uniform quantum state \cite{wang2021planar}, since only tracing out any two adjacent subsystems leaves the remaining two maximally mixed. Since AME(4,2) states do not exist \cite{ame_existence4} and are equivalent to 2-uniform states of four parties \cite{alsina_our_ame62}, a planar 2-uniform state represents the maximal amount of genuine quantum entanglement that four qubits can share over an encoding of one logical qubit.

In our simulations, we prepare the encoded logical $\ket{\bar{0}}$ state in the usual manner (that is, with one round of stabilizer measurements), before performing several cycles of stabilizer measurements, as is standard in error-detection experiments \cite{experiment1,experiment2,experiment3}. We check the \emph{logical success rate} and \emph{tripartite mutual information} on a range of one to ten cycles; our reason for this is to evaluate the time dependence of our metrics and the potential for larger-depth quantum-circuit experiments in the selected near-term spin-qubit devices. 

%--------------------------------------------%
\subsection{Entropic Measures in Monitored Quantum Circuits} \label{section:monitored_quantum_circs}

The \emph{Von Neumann entropy} (VNE) is a fundamental measure of entanglement in quantum information theory \cite{nielsenchuang}, and takes the form

\begin{equation}
\mathcal{S}(\rho) = -\text{Tr}\big[ \rho\log{\rho} \big]~,
\end{equation}

where all logarithms here are natural bases, and $\rho$ represents a density matrix.

The VNE plays an essential role in assessing the amount of quantum entanglement present in a state. By taking the partial trace of a density matrix, the resulting reduced density matrix's VNE (also known as the \emph{entanglement entropy} of the remaining subsystems) indicates whether non-trivial quantum entanglement is present; if the result is non-zero, then the two bipartitions are in fact entangled. The entanglement entropy is thus bounded as $0 \leq \mathcal{S}(\rho_{s}) \leq \log{n}$, where $s$ represents the set of remaining subsystems after a partial trace operation.  

In the study of \emph{monitored quantum circuits} (i.e., quantum circuits that are subject to frequent projective measurements) and \emph{measurement-induced phase transitions} (MIPT), it is common to measure the \emph{tripartite mutual information} \cite{tripartite1,tripartite2}, which is defined as 

\begin{equation}
\begin{split}
\mathcal{I}_{3}(A:B:C) = \mathcal{S}_{A} + \mathcal{S}_{B} + \mathcal{S}_{C} + \mathcal{S}_{ABC} - \\ \mathcal{S}_{AB} - \mathcal{S}_{BC} - \mathcal{S}_{AC}~,
\end{split}
\end{equation}

where, for example, $\mathcal{S}_{ABC}$ represents the joint $A \cup B \cup C$ subsystem. If $\mathcal{I}_{3} < 0$, then this behavior is generally ascribed to \emph{genuine multipartite entanglement} within the \emph{volume-law phase}, and is commonly considered as a \emph{quantum error correction (QEC) phase} of a quantum circuit \cite{qec_phase,qec_phase2,qec_phase3}, due to the proliferation of long-range entanglement present between subsystems (this is apparent since in a system with long-range entanglement, the total entanglement entropy for subsystems $AB$, $BC$, $AC$, and $ABC$ must exceed that of subsystems $A$, $B$, and $C$). $\mathcal{I}_{3} = 0$ is known to indicate strictly bipartite correlations among the residual subsystems, and $\mathcal{I}_{3} > 0$ can be associated with only classical correlations in the \emph{area-law phase} \cite{eisert2008area}, wherein the individual subsystems begin to behave independently of each other. 

One may ask why we do not employ a simpler analysis for estimating multipartite entanglement, such as the \emph{bipartite mutual information} \cite{bipartite1,bipartite2} or the scaling of entanglement entropy \cite{vne_scaling}. The goal of the present work is to assess how well quantum entanglement is distributed across a device, given connectivity, size, and shuttling constraints via compilation options. This setup naturally excludes the scaling of entanglement entropy, as such an analysis requires the evaluation of sequentially larger devices and circuits, in order to check if an MIPT is present; that is not the present goal of our work. Moreover, the bipartite mutual information has been used to confirm the existence of an MIPT via bipartite quantum correlations; in our case, however, we wish to check and affirm how well \emph{genuine multipartite entanglement} (i.e., multipartite entanglement that is shared among all parties, not just in a bipartite manner) exists after performing the circuits in question. As such, our analysis tools are well-calibrated for the task at hand.

Other measures of quantum correlations such as \emph{quantum discord}, \emph{entanglement robustness} \cite{discord, vidal1999robustness}, or \emph{concurrence} \cite{daniel_miller} exist; however, in this work, we opt to understand how well a certain spin-qubit architecture is able to generate quantum entanglement and stay inside of the volume-law (QEC) phase of a quantum circuit's phase diagram. In this sense, both quantum discord and entanglement robustness are not fitting, as the former concerns subtleties of quantum correlations that may not immediately pertain to entanglement, and the latter is related to entanglement properties of mixed states (which we effectively do not treat here). As for the concurrence, this entanglement monotone is related to bipartite entanglement correlations, which does not coincide with the goals of quantifying $n$-qubit entanglement correlations on a device architecture.

%--------------------------------------------%
\subsection{Estimated Success Probability} \label{section:esp_background}

% Also, we note that the concrete calculation of \cref{eq:ESP} is closely matched to the rest of the error models used in this work, ensuring compatibility

A standard method for evaluating an architecture's performance in the software regime is known as the \emph{estimated success probability} (ESP) \cite{SpinQ,ArtA,schmid2024computational,quetschlich2022predicting}. Although normally the ESP constitutes a simple multiplication of gate fidelities, we have provided a modification in this work, which can be expressed as:

\begin{equation}
ESP = \bigg[\prod_{k}\prod_{i}{F_{i,k}}\bigg] \cdot e^{-t/T_{2}} ~,
\label{eq:ESP}
\end{equation}

where $k$ represents the $k$th time step in the circuit execution, $i$ the $i$th gate in the $k$th time step and $F$ is the fidelity of the corresponding gates. The last term, inspired by \cite{helsen2018quantum} and utilized in \cite{ArtA}, introduces \emph{decoherence-induced errors}, which represent the probability of qubits staying coherent during the execution of the circuit. Here, $T_{2}$ and $t$ represent the decoherence time and the circuit-time duration, respectively. 

The original ESP model was introduced with only the first term, representing the worst-case scenario to execute a sequence of quantum gates successfully \cite{nishio2020extracting, murali2019noise} on a specific architecture. Unlike other models, ESP does not explicitly account for phase additions or subtractions by gates or noise. Instead, it simplifies the representation by using a single fidelity value that encapsulates the overall operational quality of each gate. This makes ESP the least costly to calculate among the four metrics that we have introduced, as it only scales linearly in computational complexity with the number of gates contained in a circuit. It does not suffer from the exponential growth of the Hilbert space with the increasing size and depth of a circuit and the accompanying device. Despite the simplicity of this first ESP version, some limited information on the circuit and the underlying device can still be derived indirectly. For instance, one architecture may possess a higher degree of connectivity than another; in this way, the former can achieve a reduced gate overhead when compared with the latter, thus scoring a higher ESP due to the necessity of fewer added gates.

From there, any modification can account for more information about the device's architecture. Taking \cref{eq:ESP} as an example, the parallelization capabilities of the device are indirectly reflected in $t$. More succinctly, when one device enables more gate executions within the same time frame as another, the former will exhibit a lower decoherence error rate than the latter. Another modification could add a term to model \emph{crosstalk} —a common issue in realistic device platforms \cite{sarovar2020detectingcrosstalk}. In fact, such a version has been successfully used as a figure of merit for spin-qubit architectural explorations in \cite{ArtA}. For instance, this modified ESP can reveal all kinds of architectural trade-offs. For example, maximizing the parallelization capabilities and/or qubit connectivity can incur a worse ESP result than anticipated due to increased crosstalk from more frequent close proximity interactions and more shared material components. 

One should also note that ESP can easily approach zero, especially including the modifications, as it exponentially drops with an increasing number of gates. Although such low values do not have any physical meaning after a certain small number, ESP still remains a reliable way to rank architectures provided \textit{numeric underflows}\footnote{\textit{Numeric underflows} can happen when there is a loss of accuracy in numerical calculations if ESP becomes smaller than the smallest positive representable value in a programming language's floating-point arithmetic.} are avoided. It is important to recognize that ESP is not inherently random, regardless of how small its value becomes. For example, even minor differences between two architectures, such as a single gate variation, will be consistently and reliably captured in their respective ESPs.

Overall, ESP and its modifications do not aim to absolutely predict but rather to establish a relative hierarchy of differentiation between architectures with low computational costs. However, the ESP, in and of itself, differs from the other three metrics, which can be summarized as follows. Firstly, and perhaps most importantly, ESP does not take into account notions of entanglement as in traditional quantum information-theoretic measures \cite{nielsenchuang,wilde2013quantum}. Considering that the goal of our work is to scrutinize near-term spin-qubit architectures from the standpoint of genuine multipartite entanglement, ESP stands as a cheap alternative, but it still holds its value due to its strong correlation with the other metrics, as shown in \cref{section:results}. Finally, and as explained before, no explicit parameters related to the device's architectural properties (such as the parallelization capabilities or connectivity) are inherently factored into the calculation of ESP; instead, the effects of design features are reflected in the result indirectly.

%--------------------------------------------%
\subsection{Quantum Compilation} \label{section:compilation_background}

Up until this point, we have not described our work as it relates to \emph{quantum compilation}; indeed, this is an important factor when aiming to realize a quantum algorithm in circuit form on a quantum computer. Simply put, quantum compilation constitutes the various adjustments at the level of software to a quantum circuit in order to prepare and execute it on a quantum device \cite{qiskit2024,cirq,pyquil,SpinQ}; this includes all of the subprocesses affecting a hardware-agnostic circuit, transforming it into a hardware-compatible version. Each of these procedures aims to maximize the success rate of the quantum circuit by utilizing optimization-based algorithms that take close account of the hardware’s constraints \cite{li2019tackling,steinberg2024lightcone}. Typically, these steps are subdivided into several general stages, which consist of the following, without regard to a particular ordering: a) \emph{elementary gate-set decomposition}, in which a quantum circuit is translated and simplified as much as possible in the quantum device's native gate set; b) \emph{scheduling}, wherein the logical time ordering of the circuit is considered, as well as the parallelism of gate operations and the shortening of circuit depth, among other factors; c) \emph{initial placement} (also known as \emph{qubit assignment} or \emph{initial assignment}), which assigns qubits initially from the circuit to the device; and d) \emph{qubit routing}, wherein qubits are brought in close proximity (usually adjacent) with the minimal use of a hardware’s native communication method in order to facilitate two-qubit gate interactions of the quantum circuit. 

One of the central subproblems in quantum compilation involves solving the \emph{quantum circuit mapping problem} (QCMP). The QCMP can be defined as both the initial placement and qubit routing steps of compilation combined, and as explained before, is paramount to the circuit's success rate \cite{kusyk2021survey,10313857}, especially in the \emph{noisy intermediate-scale quantum} (NISQ) and \emph{early fault-tolerant} (eFT) eras, where both QPUs and quantum circuits are relatively small, and tradeoffs exist regarding the degree to which QEC and fault-tolerance techniques are utilized \cite{katabarwa2024earlyFT,preskill2018quantum}. Lower bounds are known for the QCMP in terms of the number of \emph{SWAP} gates needed to realize a circuit on a given quantum processor with finite connectivity \cite{steinberg2024lightcone}. However, in this work, instead of scrutinizing the number of SWAP gates added by a compiler, we analyze the number of \emph{shuttles} needed, as it is known that spin-qubit quantum devices can accommodate such hardware-level operations, thereby incurring a smaller resource cost than SWAP gates in superconducting processors over sufficient length scales \cite{taylor_2005}.

\begin{figure*}
\centering
\includegraphics[width=0.95\textwidth]{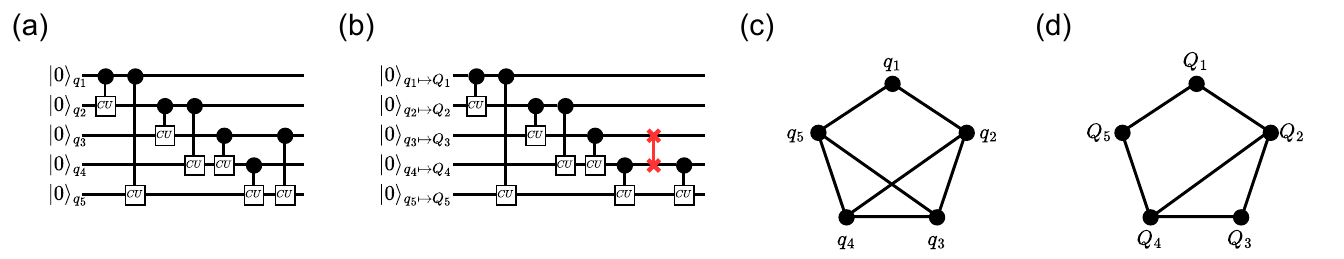}
\caption{A small example of \emph{quantum circuit mapping}; gates depicted are general controlled-unitary operations ($CU$). \emph{SWAP} gates are shown in red. (a) An algorithm in circuit form; the graph-theoretic representation of all two-qubit gates, known as the \emph{interaction graph} (IG), is shown in (c). The task of quantum circuit mapping lies in assigning qubits from a quantum circuit to the physical qubits of a device, and subsequently reorganizing physical qubits such that all two-qubit interactions are performed in an efficient manner. (b) The mapped circuit; as the hypothetical architecture in (d), represented as a \emph{connectivity graph} (CG), cannot accommodate all of the two-qubit gates represented in (c), we therefore must perform a SWAP gate.}
\label{fig:simple_quantum_compiling_example}
\end{figure*}

\cref{fig:simple_quantum_compiling_example} describes an example solution to the QCMP. (a) and (c) depict the hardware-agnostic circuit and graph-theoretic representations of the two-qubit gates utilized in an algorithm; this representation is known in the literature as the \emph{interaction graph} (IG) \cite{SpinQ}. In (b), a possible solution is presented for realizing the circuit on a quantum computer with connectivity shown as in (d); this graph is known as the \emph{connectivity graph} (CG). We can assign qubits from the algorithm to physical hardware qubits; however, as no connection on (d) exists in order to account for the two-qubit gate $CU_{q_{3},q_{5}}$ required, one solution is to add a SWAP gate between $Q_{3}$ and $Q_{4}$ such that the final two-qubit interaction can be performed. 

In order to evaluate the near-term spin-qubit architectures in our study, we map the quantum circuits presented in \cref{section:mme_qec} onto each architecture through a compilation framework. The dedicated compilation framework for spin-qubit technologies, \emph{SpinQ} \cite{SpinQ}, was utilized in order to compile circuits for our simulations. As explained in \cref{section:spin_qubit_archs}, all circuits considered have been expressed in the hardware's native gates; therefore, the \emph{elementary gate-set decomposition} stage can be skipped, and we can instead concentrate on \emph{initial placement}. As mentioned above, initial placement denotes the process of assigning qubits from an architecture-agnostic circuit to the physical qubits of the device; spin-qubit architectures are no exception to this rule. An exact solution to the QCMP is found when the IG is (sub-graph) isomorphic to the CG, inducing no extra gate overhead \cite{steinberg2024lightcone}. In practice, however, there are typically no exact initial placement solutions where all two-qubit gates can be satisfied without qubit movements; this is principally due to the bilinear nature of the considered connectivity graphs, in combination with the number and order of two-qubit gates in the featured circuits. To the best of our knowledge, no specialized initial placement algorithm has been developed for spin-qubit devices that optimizes for the intricacies of \emph{shuttling operations}. In order to compensate, we incorporated the widely used SABRE algorithm \cite{li2019tackling} into SpinQ \cite{SpinQ} for initial placement purposes.

The initial placement algorithm in SABRE works by generating a random qubit placement followed by the utilization of the \textit{SWAP-based heuristic search} routing algorithm \cite{li2019tackling} from which the resulting final qubit placement is used as the placement for the reverse circuit. This final placement derived from the reverse circuit can serve as an optimized initial placement for the original circuit. This refined initial mapping is of higher quality, as it incorporates comprehensive information about all gates and qubit interactions in the circuit. Because SABRE's placement quality heavily relies on a SWAP-based routing algorithm, fundamentally, it cannot be optimized for \emph{shuttle-based routing} \cite{beSnake}. This complication signifies a fundamental unpredictability in the initial placement quality, as a higher number of random placement trials does not necessarily imply that better solutions can be found for spin-qubit devices. 

After settling on an initial placement, we employ the \emph{beSnake} \cite{beSnake} spin-qubit routing algorithm in order to handle all two-qubit gates in a circuit, in addition to shuttling qubits towards measurement sites for readout. In the architecture CGs that we assess, we assume that shuttling is allowed bidirectionally between any two directly connected nodes and, moreover, that two-qubit gates between any adjacent qubits in the CG can be performed. In this vein, beSnake can be optimized to exploit the imposed shuttling constraints when the movement of one qubit is blocked by others. As outlined in \cite{beSnake}, certain scenarios may arise where shuttling blockages occur. In such cases, specific qubit(s) may obstruct the movement of other qubits along a shortest path. To address these blockages, dedicated mechanisms are employed. These mechanisms have been revisited, ensuring all blockages are resolved exclusively through shuttle operations.

%--------------------------------------------%
\subsection{Near-term Spin-Qubit Architectures} \label{section:spin_qubit_archs}

\begin{figure}
\centering
\begin{subfigure}[b]{0.35\textwidth}
\centering
\caption{ }
\includegraphics[width=\textwidth]{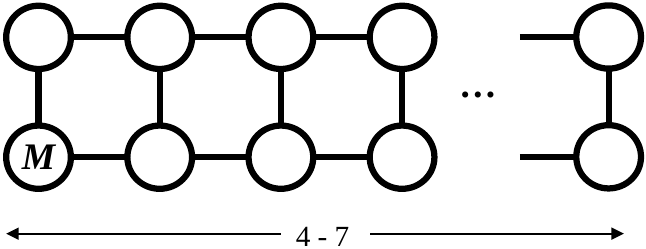}
\label{fig:connectivities1}
\end{subfigure}
\hfill
\begin{subfigure}[b]{0.35\textwidth}
\centering
\caption{ }
\includegraphics[width=\textwidth]{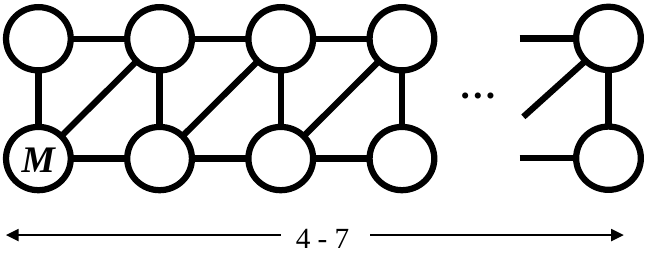}
\label{fig:connectivities2}
\end{subfigure}
\hfill
\begin{subfigure}[b]{0.35\textwidth}
\centering
\caption{ }
\includegraphics[width=\textwidth]{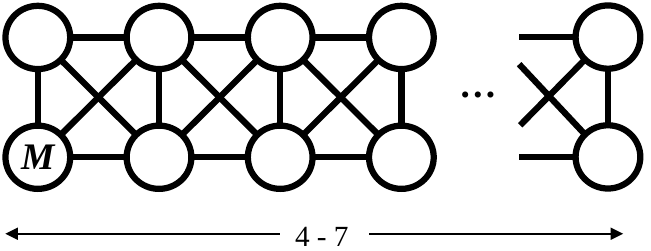}
\label{fig:connectivities3}
\end{subfigure}
\caption{Connectivity graphs (CG) of possible spin-qubit bilinear-array architectures considered in this work; we label each CG as 1, 2, and 3, in (a), (b), and (c), respectively. We analyze the entanglement properties of quantum circuits generating an AME state and a small quantum error detection code; each circuit utilizes seven qubits in total. In spin-qubit architectures, one typically does fill the entire lattice with qubits, although this is not necessary in general. As such, we test the capabilities of three different lattice connectivities shown in (a), (b), and (c) for seven qubits used in quantum circuits. We also increment the size of each bilinear array from $2 \times 4$ to $2 \times 7$ quantum dots. In all twelve hypothetical architectures, readout can be performed at the bottom-left quantum dot by offsite sensory components.} 
\label{fig:connectivities}
\end{figure}

Spin-qubit technologies possess distinctive physical properties that position them as a promising candidate for scalable quantum computing systems. One of the most notable advantages of semiconductor spin qubits is their extraordinarily small size — up to a thousand times smaller than other qubit technologies \cite{vandersypen2017interfacing}. This compactness is complemented by extensive experience in the semiconductor industry, which supports their development \cite{zwerver2022qubits, steinacker2024300, george202412}. The key component in spin-qubit technologies is the \emph{quantum dot}, which confines an electron or a hole, thus forming a physical qubit \cite{burkard2023semiconductor,hanson2007spins}. Control of a spin-qubit is achieved electromagnetically through carefully positioned gate electrodes surrounding the quantum dot. These electrodes facilitate single- and two-qubit operations (in addition to qubit \emph{shuttling}), executed via precise pulse sequences in systems with multiple quantum dots; such systems have been extensively explored in one-dimensional arrays \cite{george202412,watson2018programmable}.

More particularly, spin qubits in gate-defined quantum dots are architecturally interesting as they may be implemented in dense, highly-connected 2D quantum dot arrays \cite{veldhorst2017silicon, li2018crossbar, hill2015surface} or as sparse, low-connectivity registers connected via coherent long-distance links of multiple qubit modules \cite{boter2022spiderweb, kunne2024spinbus, taylor_2005, buonacorsi2019network}. Within the diverse range of architectural designs, various approaches have been developed to support surface code quantum error correction, each implemented in distinct ways. For the purpose of advancing quantum processor development, it is crucial to evaluate which architectures are most effective based on quantum information-theoretic principles rather than solely on experimental practicality.

While the benefits of semiconductor spin-qubit quantum processors over other technological counterparts are notable, it is known that fabrication challenges emerge in scaling them, particularly in two dimensions \cite{vandersypen2017interfacing,burkard2023semiconductor,bluhm2019semiconductor,de2023silicon,de2021materials}. On the positive side, there have been significant efforts \cite{li2018crossbar,franke2019rent,paquelet2020multiplexed,pauka2019cryogenic,veldhorst2017silicon,buonacorsi2019network,kunne2024spinbus} to tackle these challenges and scale them in higher dimensions. It is forecasted that \emph{bilinear arrays} are especially promising for near-term fabrication and experiments due to their amenity to current technological capabilities, making them more realistic to build \cite{xue2022performance,siegel2024towards}. Thus, in this work, we consider bilinear arrays of various sizes and connectivities as prospective candidates for near-term spin-qubit devices.

Our goal is to evaluate the merits and drawbacks of these devices by considering chiefly: a) the expected quantumness in their qubit-qubit entanglement correlations; b) anticipated logical success rates and ESP (Estimated Success Probability) resulting from encoding a small error-detection code; and c) the degree to which genuine multipartite quantum entanglement is generated and maintained on the device. In \cref{fig:connectivities}, we illustrate these bilinear arrays through their connectivity graphs (CG). In a CG, each node (circle) represents a site of the grid, and each edge connecting the nodes corresponds to the possibility of interaction between the two sites. The CG representing the arrays alongside further hardware characteristics comprises the architectures treated in this work. Additionally, readout can be performed at the bottom-left quantum dot by offsite sensory components. Usually, in experimental devices, it is possible to use multiple measurement sensors. However, we opted to simplify the compilation process by standardizing the decision of which site should be used for all readouts. Additionally, this approach allowed us to focus exclusively on comparing the sizes and connectivity graphs of the circuits, as introducing an extra sensor could create an unfair advantage for certain configurations, potentially skewing the comparison by introducing size-specific benefits.

In a quintessential semiconductor spin-qubit device, it is optimistic to consider the following quantum device-specific properties at present \cite{camenzind2022hole,hendrickx2021four,chatterjee2021semiconductor,RevModPhys.85.961,PhysRevA.57.120,vandersypen2017interfacing,veldhorst2015two,zajac2015reconfigurable,watson2018programmable}:

\begin{itemize}
\item A coherence $T_{2}^*$ time of \SI{20}{\micro\second} \cite{steinacker2024300}; 
\item Single-qubit \cite{yoneda2018quantum} and shuttle durations of \SI{100}{\nano\second} \cite{yoneda2021coherent, van_Riggelen_Doelman_2024}; 
\item Two-qubit gate times of \SI{150}{\nano\second} \cite{xue2022quantum, noiri2022fast}; and
\item Measurement times of \SI{5}{\micro\second} \cite{Takeda_2024}.
\end{itemize}

In such optimistic models, \emph{thermal relaxation} (i.e., the $T_{1}$ time) usually ranges between \SI{100}{\milli\second} to a few seconds. Regarding gate fidelity, we assume an average of 99.99\% for single-qubit gates and nearest-neighbor shuttles, 99.90\% for two-qubit gates, and readout \cite{philips2022universal,xue2022quantum,noiri2022fast}. Array initialization and qubit resets can also contribute to errors; however, for simplicity in our analysis, we assume that these values are ideal.

To formulate the gate set for our quantum circuits, we ensured that all circuits were expressed using a predefined gate set. Assuming this gate set is not natively supported by the underlying hardware, further decomposition may be required. However, since our focus lies in architectural comparisons rather than in absolute performance, such a tactic will not change the nature of our simulations, since all circuits or produced compilation overhead will experience the same relative change. For these reasons, we selected the gate set $\{H, CX, CY, CZ\}$, which facilitates constructing quantum circuits with a minimal number of gates, highlighting the adaptability and flexibility of our approach to meet various requirements.

It is additionally assumed that no gates can be performed in parallel; there are two main reasons for this. Firstly, it is known that pernicious and notoriously troublesome \emph{crosstalk effects} are known to arise during parallelization in spin-qubit experiments \cite{heinz2021crosstalk}. Secondly, current generation spin-qubit experiments do not incorporate gate-parallelization techniques, although this may change in the future \cite{patomaki2024pipeline}. For these reasons, our first two error models are crosstalk-free. However, as a proof of concept for the utility of our work, we showcase a basic implementation of crosstalk-induced errors in \cref{section:cl_crosstalk_sims}, and describe how the current error models can be extended. In accordance with \cite{sarovar2020detectingcrosstalk}, we define a CPHASE$(\zeta)$ gate as 

\begin{equation}
\text{CPHASE}(\zeta) = 
\begin{bmatrix}
1 & 0 & 0 & 0\\
0 & 1 & 0 & 0\\
0 & 0 & 1 & 0\\
0 & 0 & 0 & e^{i\zeta}
\end{bmatrix}~,
\end{equation}

where $\zeta \in [0, 2\pi]$. The actual error model involving crosstalk will be defined more deeply in \cref{section:cl_crosstalk_sims}.

As a final note, the routing algorithm beSnake was modified such that when a measurement needs to take place, routing can commence in order to move a qubit toward the measurement zone of a bilinear array. We have enabled the routing optimization feature where a path-selection heuristic evaluates the best shortest path within a 0.05-second time limit, which is plenty for the circuit and device sizes of this study, as tested in \cite{beSnake}.

%--------------------------------------------%
\section{Results} \label{section:results}

%--------------------------------------------%
\subsection{Simulation \& Error Model} \label{section:simulation_error_model}

Results from each of the simulations realized are shown in \cref{fig:ame_results,fig:logical_qubit_ESP_results,fig:logical_success_rates_results,fig:tensor_network_results}, corresponding to: a) extrapolation of the Bell operator for the AME(6,2) state (\cref{fig:ame_results}); b) calculation of the logical success rate for the $\llbracket 4,1,2 \rrbracket$ error-detecting surface code (\cref{fig:logical_success_rates_results}); c) the ESP for circuits pertaining to the error-detecting surface code (\cref{fig:logical_qubit_ESP_results}); and d) the results from calculating the tripartite mutual information (\cref{fig:tensor_network_results}). For each data point generated, either $10^{4}$ (\cref{fig:ame_results}), $2\times10^{4}$ (\cref{fig:logical_success_rates_results,fig:logical_qubit_ESP_results}), or $2\times10^{3}$ (\cref{fig:tensor_network_results}) Monte Carlo trials were utilized. Furthermore, we used a specific seed number on the non-deterministic aspects of our simulations for reproducibility and consistency in our comparisons.

The error models utilized in our simulations differ depending upon whether we utilized Qiskit \cite{qiskit2024} or a tensor-network (TN) simulation via the Python package quimb \cite{gray2018quimb}; as such, we name the corresponding error channels applied in our simulations as $\mathcal{E}_{\text{Q}}$ and $\mathcal{E}_{\text{TN}}$, respectively. These error models are shown graphically in \cref{fig:both_error_models}, and they both take on the following forms:

\begin{figure}
    \centering
    \includegraphics[width=0.9\columnwidth]{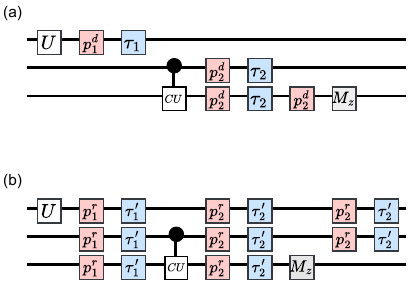}
    \caption{Both of the error models used in this work. (a) denotes the error model $\mathcal{E}_{\text{Q}}$, while (b) displays the error model $\mathcal{E}_{\text{TN}}$.}
    \label{fig:both_error_models}
\end{figure}

Error model $\mathcal{E}_{\text{Q}}$ is defined as follows:

\begin{itemize}
\item After a single-qubit gate $U$ in our model, with probability $p^{d}_{1}$ a random Pauli error (that is, an error drawn from $\{$X, Y, Z$\}$) is applied;
\item Additionally, after the same single-qubit gate $U$, with probability $\tau_{1} = \frac{1 - e^{-t/T_{2}}}{2}$, a Z error is applied, where $t$ depends on the single-qubit gate duration; 
\item After a two-qubit gate $CU$, we apply with probability $p^{d}_{2} = 10 p^{d}_{1}$ a random Pauli error (that is, an error drawn from $\{$X, Y, Z$\}$) on each of the qubits involved;
\item Moreover, after the same two-qubit gate $CU$, with probability $\tau_{2} = \frac{1 - e^{-t_{2}/T_{2}}}{2}$, a Z error is applied, where $t_{2}$ depends on the two-qubit gate duration;
\item Before each measurement $M_{z}$, with probability $p^{d}_{2}$ we apply an error drawn from $\{$X, Y, Z$\}$;
\end{itemize}

The error model $\mathcal{E}_{\text{TN}}$ is defined as follows:

\begin{itemize}
\item After a single-qubit gate $U$ in this model, with probability $p^{r}_{1}$, a random rotation gate is performed on each qubit, drawing from the set $\{ R_{x}(\phi), R_{y}(\phi), R_{z}(\phi) \}$, where $\phi = \pi / \mu$ and $\mu \in (0,1]$;  
\item In the same time step and after the same single-qubit gate $U$, with probability $\tau'_{1}$ a Z-rotation gate $R_{z}(\psi)$, where $\psi = \pi e^{t/T_{2}}$, is applied on each qubit; 
\item After a two-qubit gate $CU$, we apply with probability $p^{r}_{2} = 10 p^{r}_{1}$ a random rotation gate, drawing from the set $\{ R_{x}(\phi), R_{y}(\phi), R_{z}(\phi) \}$, where $\phi = \pi / \mu$ and $\mu \in (0,1]$ for all qubits;
\item After the same two-qubit gate $CU$, with probability $\tau'_{2} = 10\tau'_{1}$ a Z-rotation gate $R_{z}(\psi)$, where $\psi = \pi e^{t/T_{2}}$, is applied on each qubit;
\item After a measurement $M_{z}$ is performed, we apply with probability $p^{r}_{2} = 10 p^{r}_{1}$ a random rotation gate, drawing from the set $\{ R_{x}(\phi), R_{y}(\phi), R_{z}(\phi) \}$, where $\phi = \pi / \mu$ and $\mu \in (0,1]$ for all qubits;
\item Finally, after that same measurement $M_{z}$ is performed, with probability $\tau'_{2} = 10\tau'_{1}$ a Z-rotation gate $R_{z}(\psi)$, where $\psi = \pi e^{t/T_{2}}$, is applied on each qubit.
\end{itemize}

The error models $\mathcal{E}_{\text{Q}}$ and $\mathcal{E}_{\text{TN}}$ differ fundamentally in their design and application. With these choices, we aim to demonstrate the diversity of our approach by employing different error models, each offering distinct trade-offs between computational complexity and realism. Further insights are provided in detail in \cref{section:discussion}. In this context, the tensor-network simulation can be viewed as a more realistic analog for expected quantum device behavior, wherein errors are excluded after measured qubits, as these states are collapsed and reinitialized with similar \emph{fault-tolerant quantum circuit} models used in the literature \cite{bhatnagar2023low,chao2018quantum,chamberland2018flag,flag_schedule}. Conversely, the $\mathcal{E}_{\text{Q}}$ model combines the standard depolarizing model, implemented with the \texttt{depolarizing\_error()} function from Qiskit Aer library \cite{qiskit2024}, with decoherence errors cascaded using the thermal relaxation function \texttt{thermal\_relaxation\_error()}.

%--------------------------------------------%
\subsection{AME State-Generation Circuit Results} \label{section:ame_state_results}

\begin{figure*}
\centering
\begin{subfigure}[b]{0.325\textwidth}
\centering
\caption{ }
\includegraphics[width=\textwidth]{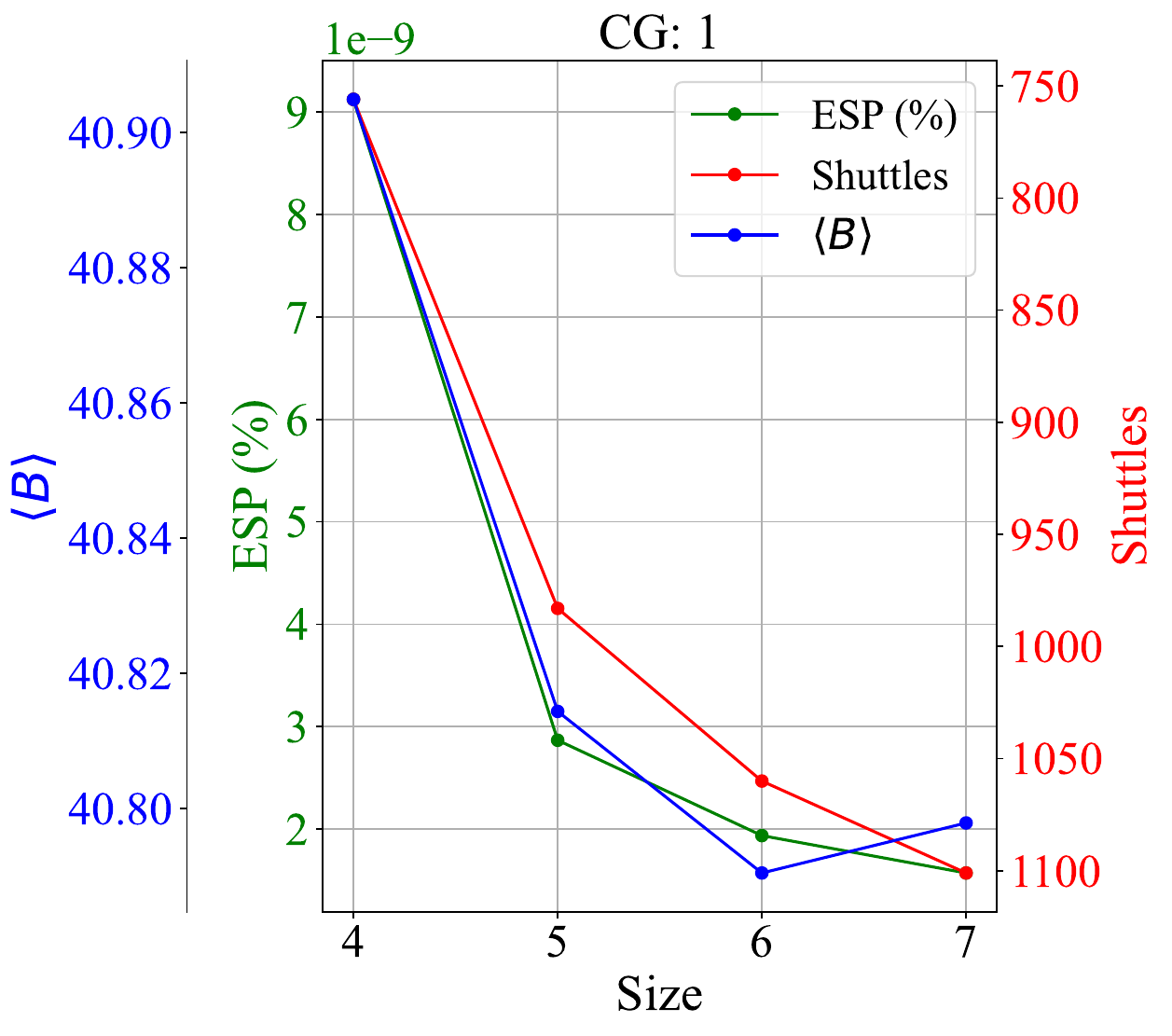}
\label{fig:ame_resultsa}
\end{subfigure}
\hfill
\begin{subfigure}[b]{0.325\textwidth}
\centering
\caption{ }
\includegraphics[width=\textwidth]{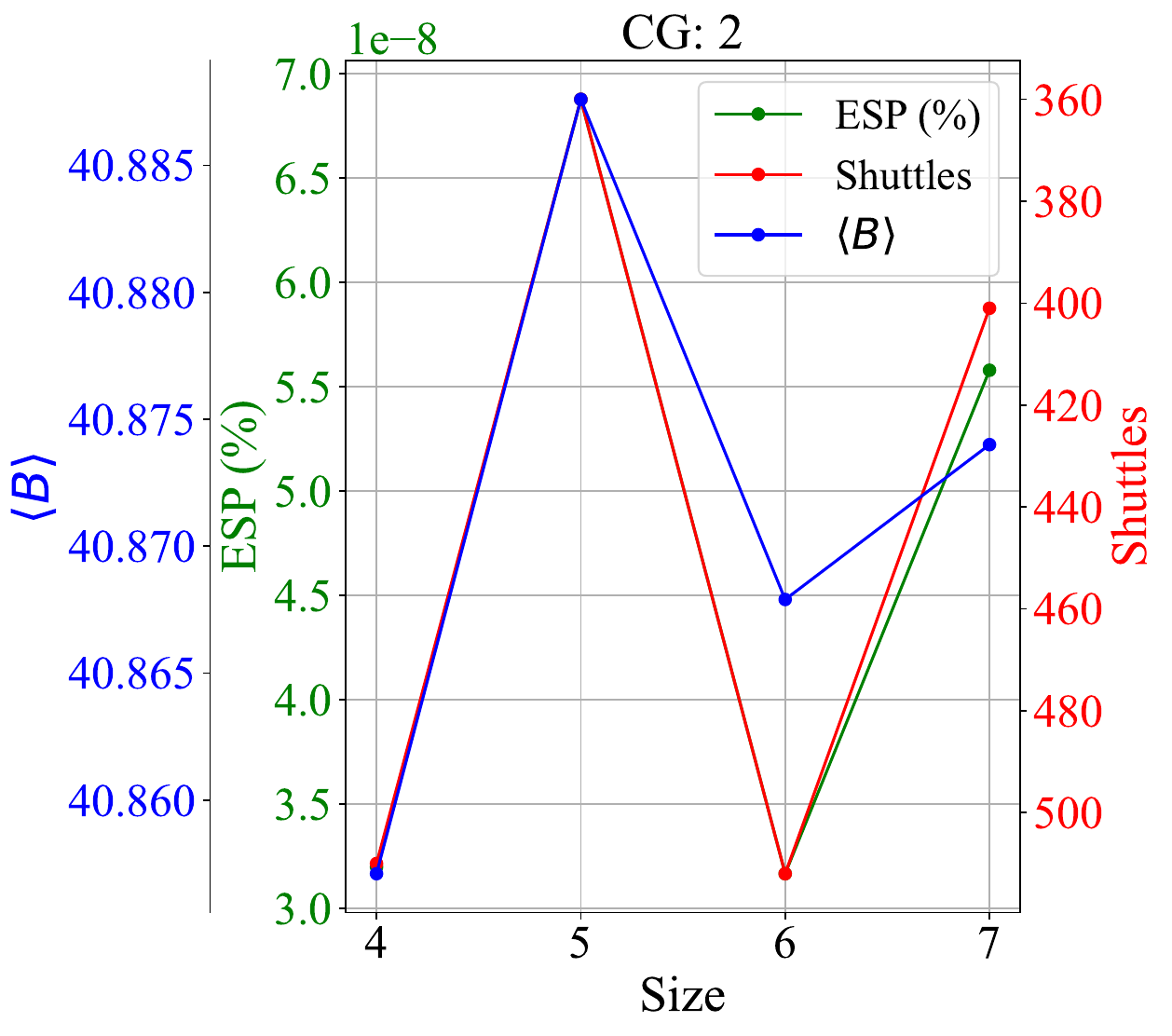}
\label{fig:ame_resultsb}
\end{subfigure}
\hfill
\begin{subfigure}[b]{0.325\textwidth}
\centering
\caption{ }
\includegraphics[width=\textwidth]{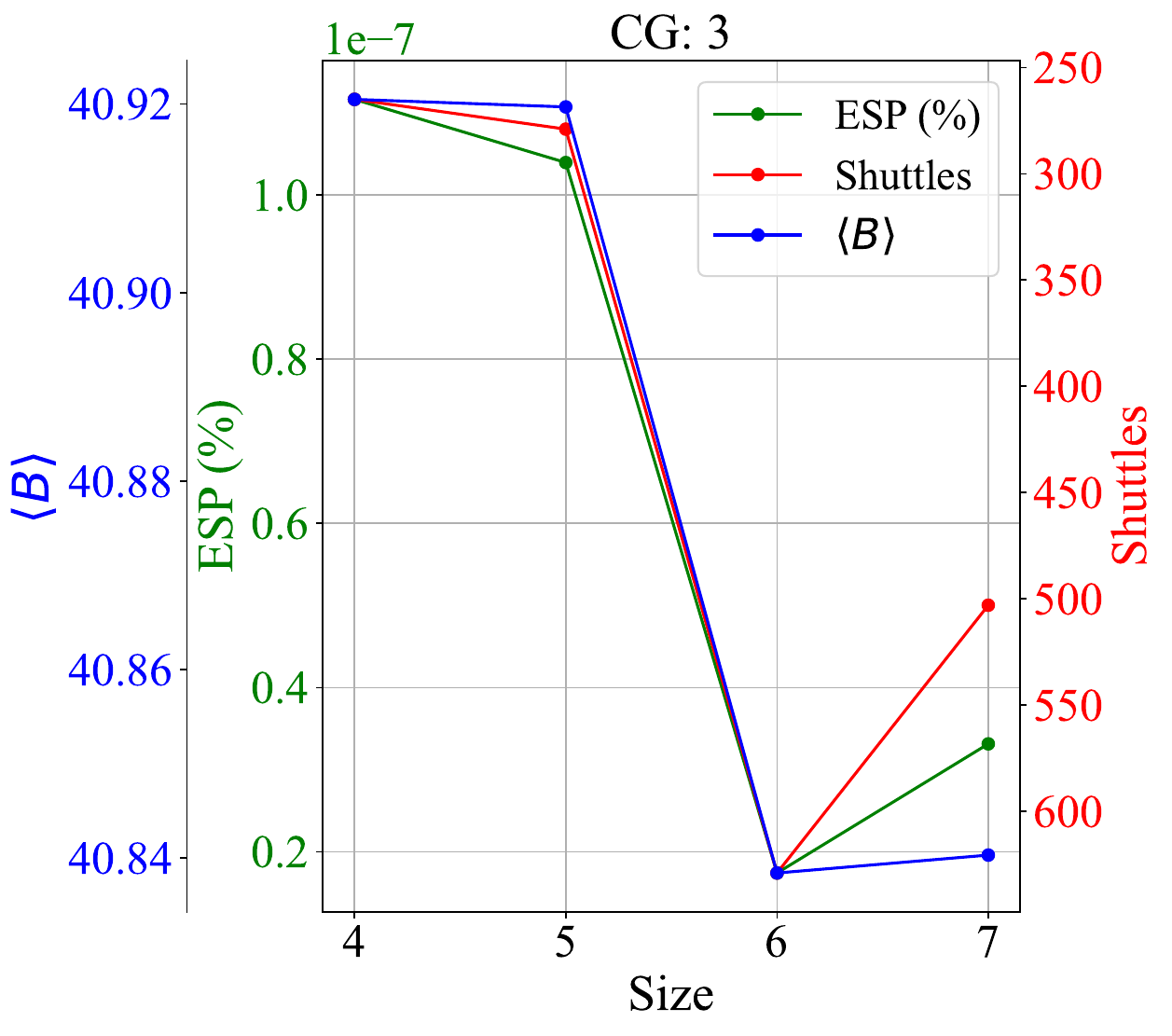}
\label{fig:ame_resultsc}
\end{subfigure}
\caption{Results for the circuit generating the AME(6,2) state. In blue, green, and red, we have measured: the average Bell operator $\langle B \rangle$; the ESP, expressed as a percentage; and the shuttle count added during the compilation process. $10^{4}$ trials were conducted for all three CGs.}
\label{fig:ame_results}
\end{figure*}

In \cref{fig:ame_results}, we have graphed the results obtained from the generating circuit for the AME(6,2) state and the subsequent average measurement of the Bell operator $\langle B \rangle$, alongside the ESP and the shuttle count added to the circuit during the compilation process. Here and in the subsequent section (\cref{section:logical_success_results}), the noise model $\mathcal{E}_{\text{Q}}$ is used. The ESP results (expressed as a percentage), the average Bell operator expectation $\langle B \rangle$, and the shuttle count have been graphed together. Subfigures (a)-(c) represent each CG from \cref{fig:connectivities}, with increasing lattice sizes from $2 \times 4$ to $2 \times 7$ quantum dot sites; as stated earlier, the net result of increasing the number of quantum dots is to reduce the density of qubits relative to unoccupied sites in the device.

In all of the figures, it is clear that the shuttle count closely correlates with the calculated $\langle B \rangle$ and ESP values. For CG$_{1}$ and CG$_{3}$, one can surmise via the expectation of  $\langle B \rangle$ that qubit-qubit correlations diminish as the size of the lattice is incremented; as a direct result, it is anticipated that the ESP drops commensurately. One exception to this trend can be seen in CG$_{2}$, wherein the metrics fluctuate on even and odd lattice sizes. One naive reason for this may be due to some extraneous effect that our simulations may not have originally accounted for; however, a more logical reason has to do with the \emph{initial placement} algorithm, SABRE, that was utilized in this work. This point will be analyzed more deeply in \cref{section:discussion}. As a final note, we additionally calculated the average (over all lattice sizes) ESP, $\langle B \rangle$, and shuttles required; these data are shown in \cref{table:esp_ame}. The overall trend for all CGs is that, as the connectivity of the device increases, the shuttle count decreases, thereby incrementing the ESP and $\langle B \rangle$ values; nevertheless, we also notice that, as the connectivity increases, the degree to which all three fields change also steadily decreases; we also touch upon this in \cref{section:discussion}. 

\begin{table}
\begin{tabular}{ |c|c|c|c| } 
\hline
CG & ESP ($10^{-8}$) & $\langle B \rangle$ & Shuttles \\
\hline 
CG$_{1}$ & 0.39 & 40.82 & 975 \\
CG$_{2}$ & 4.71 & 40.87 & 445.75 \\
CG$_{3}$ & 6.65 & 40.88 & 419 \\
\hline
\end{tabular}
\caption{Averaged ESP, $\langle B \rangle$, and shuttle operations, as reported in \cref{fig:ame_results}, over all sizes of each CG.}
\label{table:esp_ame}
\end{table}

%--------------------------------------------%
\subsection{Logical Success Rate \& ESP Results} \label{section:logical_success_results}

\begin{figure*}
\centering
\begin{subfigure}[b]{0.497\textwidth}
\centering
\caption{ }
\includegraphics[width=\textwidth]{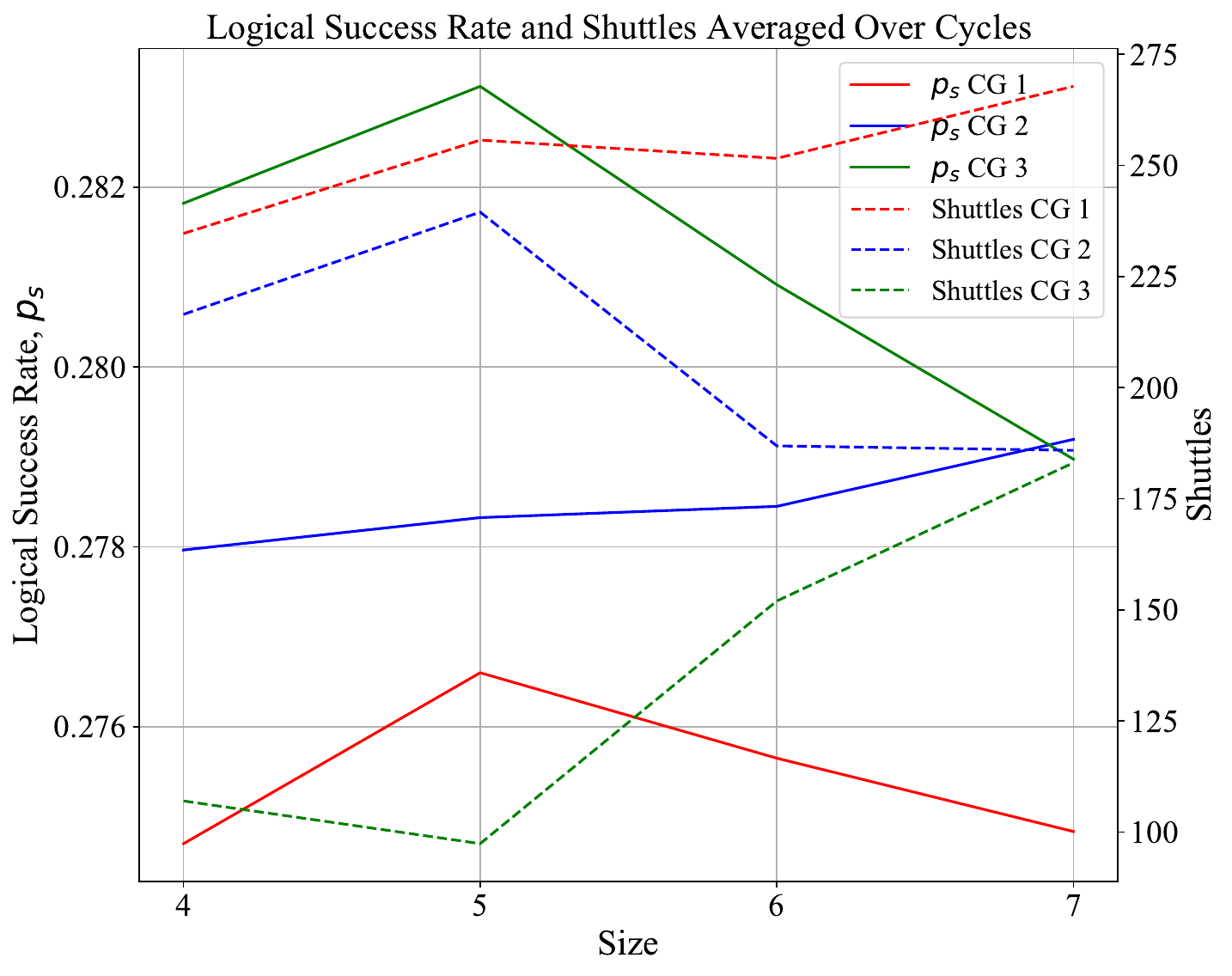}
\label{fig:logical_success_rates_resultsa}
\end{subfigure}
\hfill
\begin{subfigure}[b]{0.497\textwidth}
\centering
\caption{ }
\includegraphics[width=\textwidth]{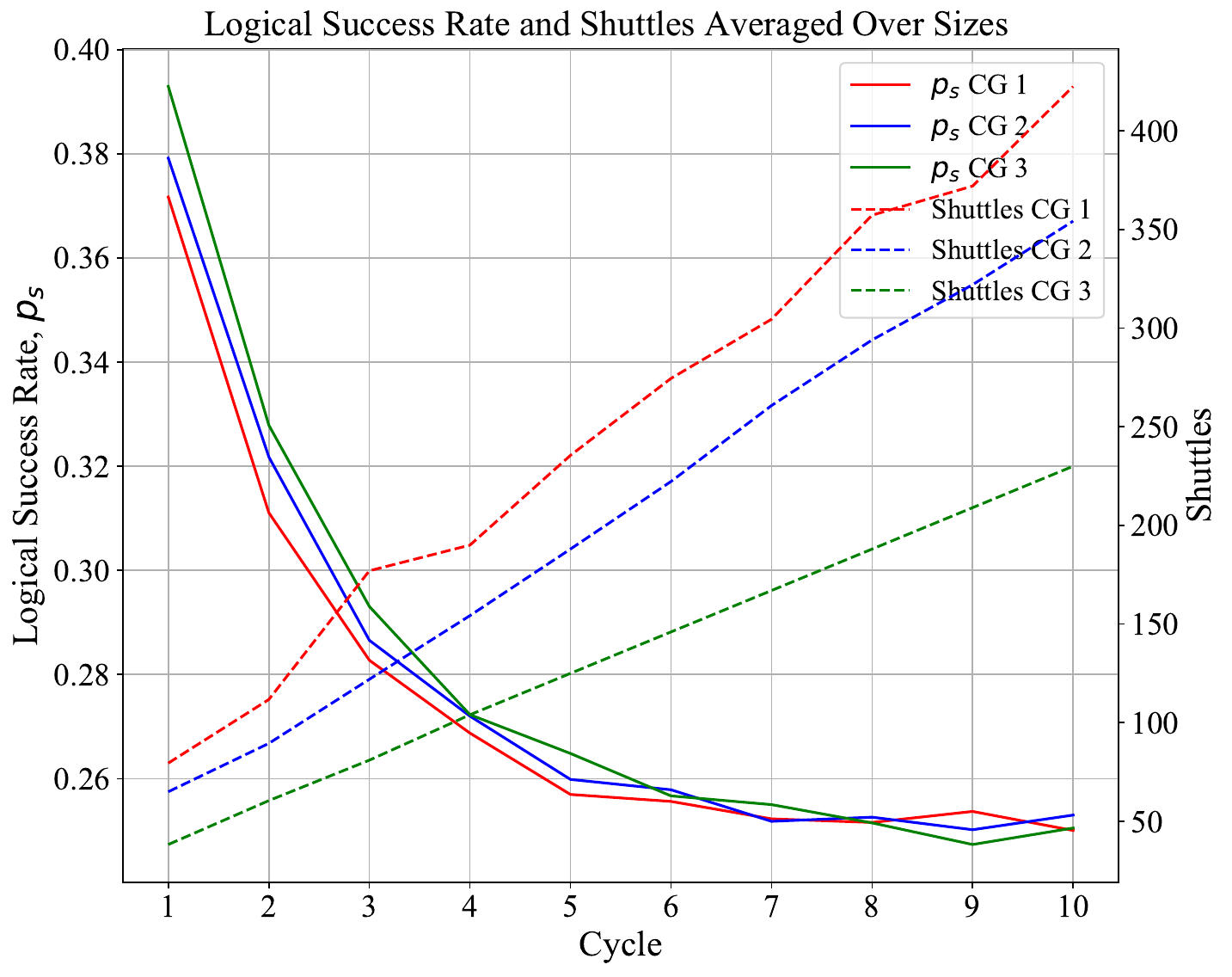}
\label{fig:logical_success_rates_resultb}
\end{subfigure}
\caption{Finalized logical success rate $p_{s}$ results for the $\llbracket 4,1,2 \rrbracket$ surface error-detecting code, under (a) cycle averaging and (b) averaging over all sizes for a particular CG, and were taken after averaging over $20,000$ trials.}
\label{fig:logical_success_rates_results}
\end{figure*}

\cref{fig:logical_success_rates_results} displays the logical success rate $p_{s}$ obtained for the $\llbracket 4,1,2 \rrbracket$ error-detecting surface code, together with the shuttles added. We have reported averages over (a) the number of cycles tested in our simulations as we increase the lattice size, and (b) over all lattice sizes as the number of cycles is incremented. As can be seen in (a), the shuttle count steadily increases for all but CG$_{2}$ (for which the shuttle count actually decreases slightly). As a result, the lattice size increases commensurately, showing a drop in the logical qubit's logical success rate. The only exception to this rule is shown for CG$_{2}$, which again exhibits an increase in logical success rate as the lattice size is increased. As for (b), averaging over lattice sizes and incrementing the number of stabilizer measurement cycles paints a different picture: initially, after one cycle, it can be seen that the logical success rate and the shuttle count inversely correlate with one another. Indeed, it is also evident that the most highly-connected CG, CG$_{3}$, exhibits a high logical success rate. However, as we raise the number of measurement cycles, at about the sixth cycle, we see that the logical success rates of all connectivity graphs are roughly indistinguishable from each other, and this trend continues as we continue to add stabilizer measurement cycles; we elaborate more on this in \cref{section:discussion}. Finally, we also report on the Pauli logical error content detected in \cref{section:appendix_logical_success_with_pauli_errors}.

\begin{figure*}
\centering
\begin{subfigure}[b]{0.497\textwidth}
\centering
\caption{ }
\includegraphics[width=\textwidth]{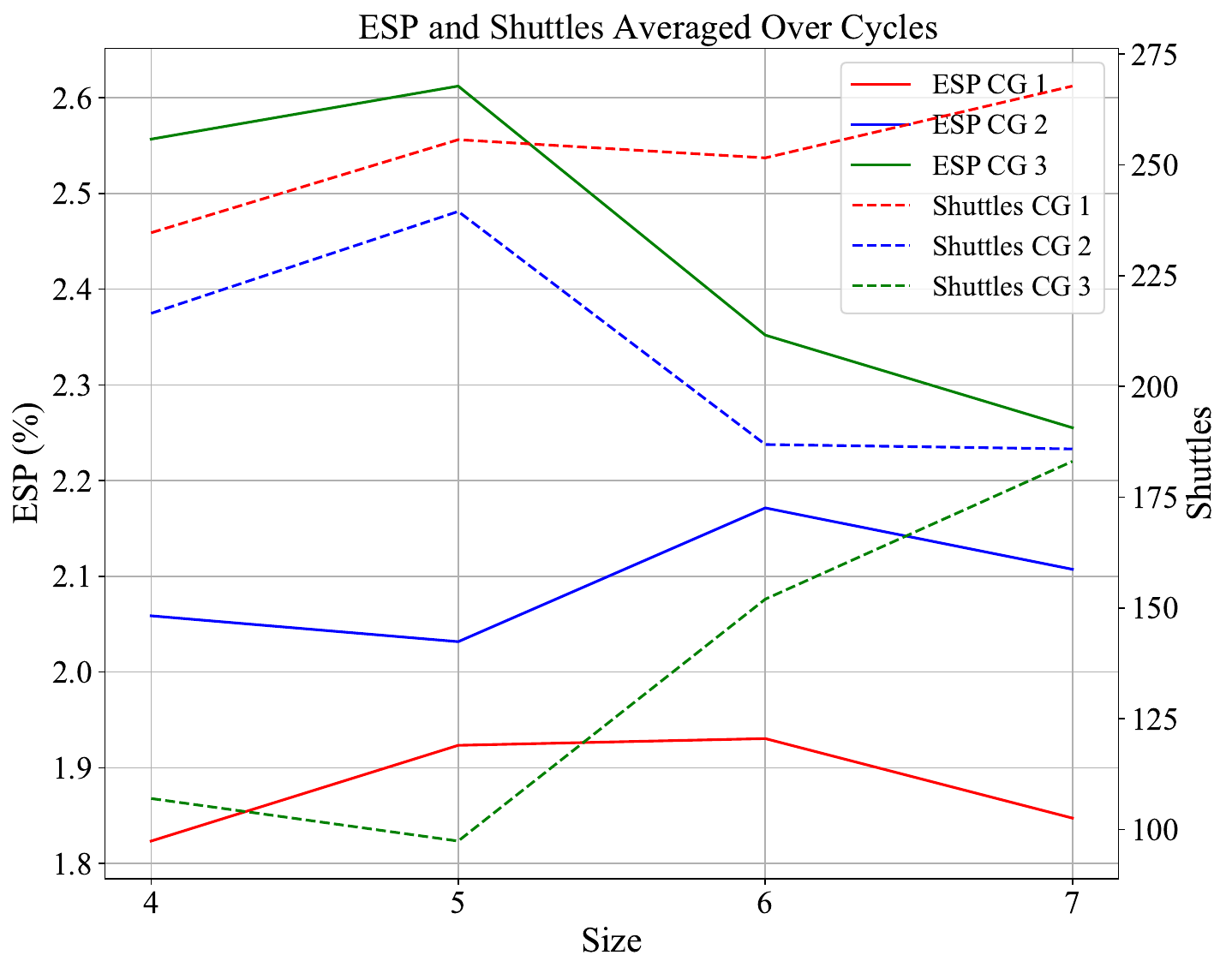}
\label{fig:logical_qubit_ESP_resultsa}
\end{subfigure}
\hfill
\begin{subfigure}[b]{0.497\textwidth}
\centering
\caption{ }
\includegraphics[width=\textwidth]{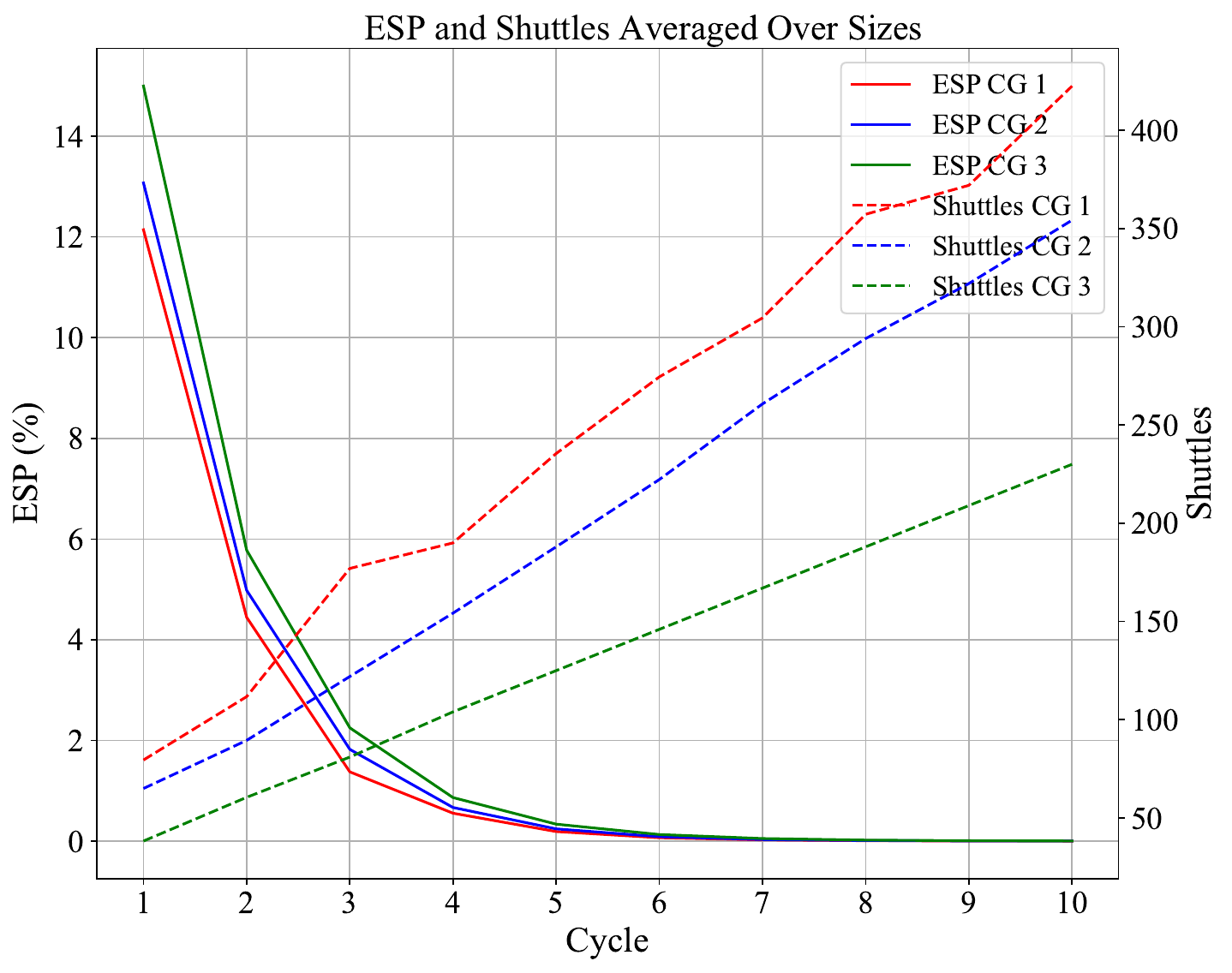}
\label{fig:logical_qubit_ESP_resultsb}
\end{subfigure}
\caption{Finalized ESP results for the $\llbracket 4,1,2 \rrbracket$ surface error-detecting code, under (a) cycle averaging and (b) averaging over all sizes for a particular CG.}    
\label{fig:logical_qubit_ESP_results}
\end{figure*}

By and large, we see in (a) and (b) of \cref{fig:logical_qubit_ESP_results} that the tendencies of both the ESP and logical success rate agree with each other, although the absolute values of both widely differ. As expected, the ESP exponentially decays with circuit depth, and converges to zero at about six or seven stabilizer measurement cycles. Again, this result is in agreement with our results in \cref{fig:logical_success_rates_results} on the logical success rate decay, and are indicative of how, in a more realistic experiment, it is expected that the logical success rate of a logical qubit converges to a specific (low) value. The reason for this convergence is related to the projective (stabilizer) measurements utilized; although errors will build up and propagate throughout a circuit as its depth is increased, projective measurements still will project the logical qubit into a logical eigenstate pertaining to one of the codestates, and in doing so, there will be a finite probability that one of these corresponds to the correct logical qubit codeword. 

In \cref{table:logical_ESP_averages}, we have calculated the averages of ESP, logical success rate, and shuttles over all cycles for each size and CG. When focusing on the average shuttles, we note a bigger reduction between CG$_{2}$ and CG$_{3}$ compared to CG$_{1}$ and CG$_{2}$. These relative differences are also reflected in the ESP and logical success rate in the table. Smaller lattice sizes can offer competitive results when compared to larger ones across all metrics displayed. These observations underscore the fact that improving connectivity or size of the device is not guaranteed to provide an improvement in the entanglement measures that we have chosen to study; we discuss this and other related observations and put them into the larger perspective in \cref{section:discussion}. 

\begin{table}
    \centering
    \begin{tabular}{|c|c|c|c|}
    \hline
        & ESP $(\%)$ & Logical Success Rate $(p_{s})$ & Shuttle Count \\
\hline
        CG$_{1}$ & 1.8812 & 0.2754 & 252.45 \\
        CG$_{2}$ & 2.0923 & 0.2785 & 207.20 \\
        CG$_{3}$ & 2.4441 & 0.2812 & 134.88 \\
        \hline
        Size 4 & 2.1463 & 0.2782 & 186.07 \\
        Size 5 & 2.1891 & 0.2793 & 197.53 \\
        Size 6 & 2.1513 & 0.2783 & 196.83 \\
        Size 7 & 2.0700 & 0.2777 & 212.27 \\
\hline
    \end{tabular}
    \caption{Averaged ESP, logical success rate, and shuttle operations, as reported in \cref{fig:logical_success_rates_results,fig:logical_qubit_ESP_results}, over all cycles of each size and CG.}
    \label{table:logical_ESP_averages}
\end{table}

%--------------------------------------------%
\subsection{More Invasive Noise Models} \label{section:cl_level_noise_sims}

\begin{figure*}
\centering
\begin{subfigure}[b]{0.497\textwidth}
\centering
\caption{ }
\includegraphics[width=\textwidth]{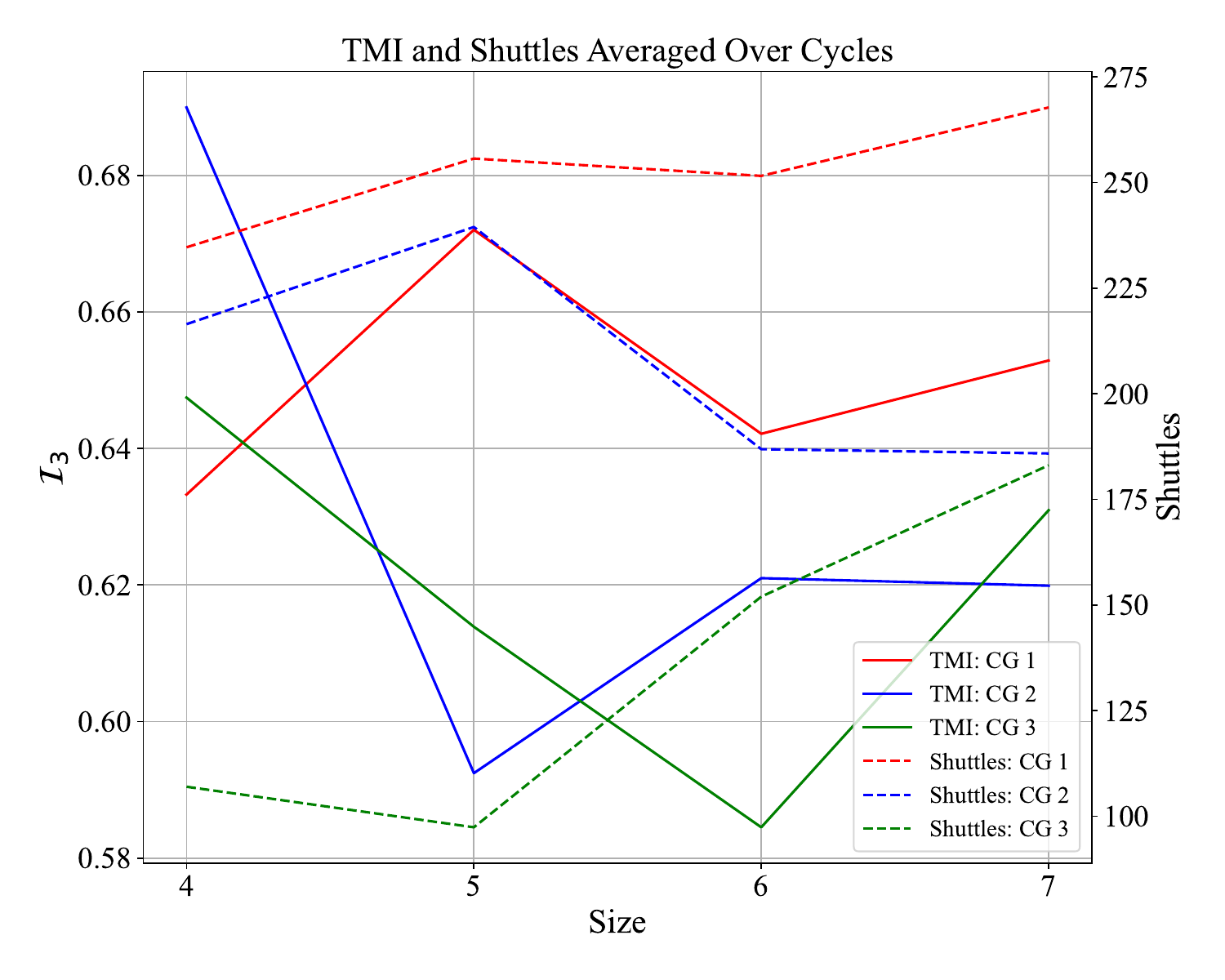}
\label{fig:tensor_network_resultsa}
\end{subfigure}
\hfill
\begin{subfigure}[b]{0.497\textwidth}
\centering
\caption{ }
\includegraphics[width=\textwidth]{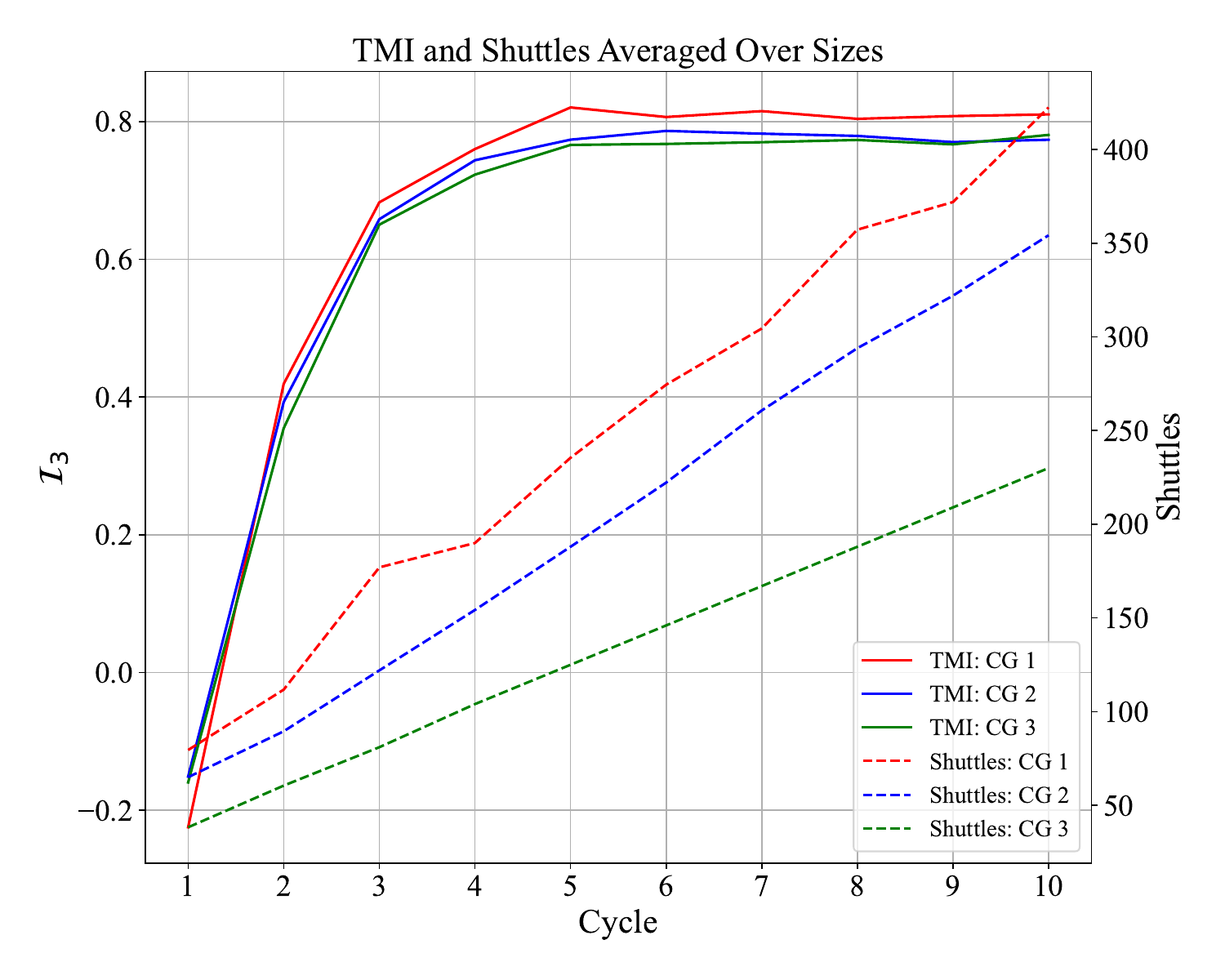}
\label{fig:tensor_network_resultsb}
\end{subfigure}
\caption{Results from the tensor network simulation of the $\llbracket 4,1,2 \rrbracket$ surface error-detecting code, under (a) cycle averaging and (b) averaging over all sizes for a particular CG. $2,000$ Monte Carlo trials were taken per data point. Tripartite mutual information is generally shown to become more positive in both (a) and (b), as additional shuttles are added to the compiled circuit; this is consistent with the $\mathcal{I}_{3}$, which is observed to leave the volume-law (QEC) phase at approximately the second cycle of stabilizer measurements, entering the area-law (disentangling) phase.}
\label{fig:tensor_network_results}
\end{figure*}

\cref{fig:tensor_network_results} depicts the tripartite mutual information $\mathcal{I}_{3}$ obtained from tensor-network simulations; these have been derived from the same compiled circuits as previously  (\cref{fig:logical_qubit_ESP_results,fig:logical_success_rates_results}). Here, the noise model $\mathcal{E}_{\text{TN}}$ is used. As $\mathcal{I}_{3}$ is a measure of the global/local distribution of quantum entanglement over subsystems, we note that, unlike previous results, a lower (potentially below zero) $\mathcal{I}_{3}$ value is desirable, as it implies a global distribution of genuine multipartite entanglement over a circuit (and therefore over the quantum device). Subfigures (a) and (b) again show the averaged $\mathcal{I}_{3}$ with respect to cycles or lattice sizes, respectively; we then opt to increment lattice sizes and the cycle number, respectively. 

In \cref{fig:tensor_network_resultsa}, we note a slightly different tendency in the $\mathcal{I}_{3}$ than in our previous simulations; namely, as the lattice size is increased for CG$_{2}$ and CG$_{3}$, at first, we observe a \emph{decrease} in the $\mathcal{I}_{3}$, and subsequently an increase (for CG$_{1}$, we notice a trend of increasing $\mathcal{I}_{3}$). The main reason for this behavior involves the observation that all of the averaged $\mathcal{I}_{3}$ values are greater than zero; that is, we can surmise that, averaged over all cycles, the tripartite mutual information for every device size is expected to stay squarely in the \emph{area-law} phase. As such, monitoring changes in the $\mathcal{I}_{3}$ is important, as this signals the changing nature of multipartite entanglement correlations in the quantum system; however, the overall character of entanglement correlations observed stay within the disentangling (area-law) phase. 

\cref{fig:tensor_network_resultsb} showcases a similar concept as in \cref{section:logical_success_results}; that is, in the first cycle, the average $\mathcal{I}_{3}$ over all lattice sizes is generally below zero, located inside the \emph{QEC phase}; here, it is presumed that \emph{volume-law} entanglement behavior dominates in the quantum system. However, as we increase the number of cycles in our simulation, we see that already in the second cycle, one can preliminarily note the presence of a \emph{measurement-induced phase transition}, as $\mathcal{I}_{3}$ crosses over from negative to positive. This information is important, as from here on out, it is known that the individual subsystems of the quantum device will begin to behave in a classically correlated manner; indeed, as we increase the number of cycles, the exact same correlation between shuttle count and $\mathcal{I}_{3}$ follows the trends of \cref{section:ame_state_results,section:logical_success_results}. Additionally, we notice that at around six or seven cycles, the $\mathcal{I}_{3}$ measured for CG$_{2}$ and CG$_{3}$ effectively converge, suggesting that, as the number of cycles is increased, the potential gain in connectivity via CG$_{3}$ matters less and less.  

%--------------------------------------------%
\subsubsection{Incorporating Crosstalk} \label{section:cl_crosstalk_sims}

\begin{figure*}
\centering
\begin{subfigure}[b]{0.497\textwidth}
\centering
\caption{ }
\includegraphics[width=\textwidth]{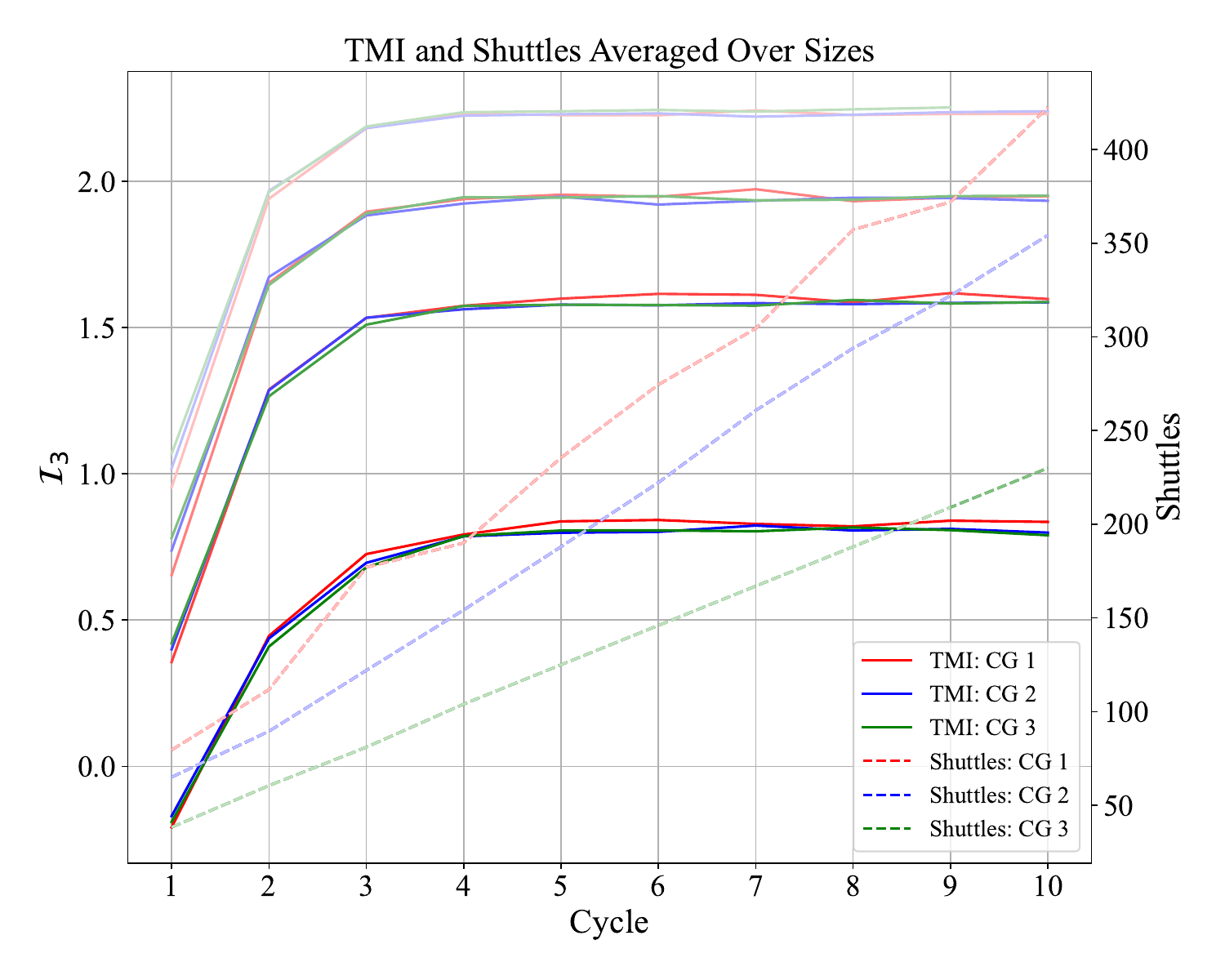}
\label{fig:crosstalka}
\end{subfigure}
\hfill
\begin{subfigure}[b]{0.497\textwidth}
\centering
\caption{ }
\includegraphics[width=\textwidth]{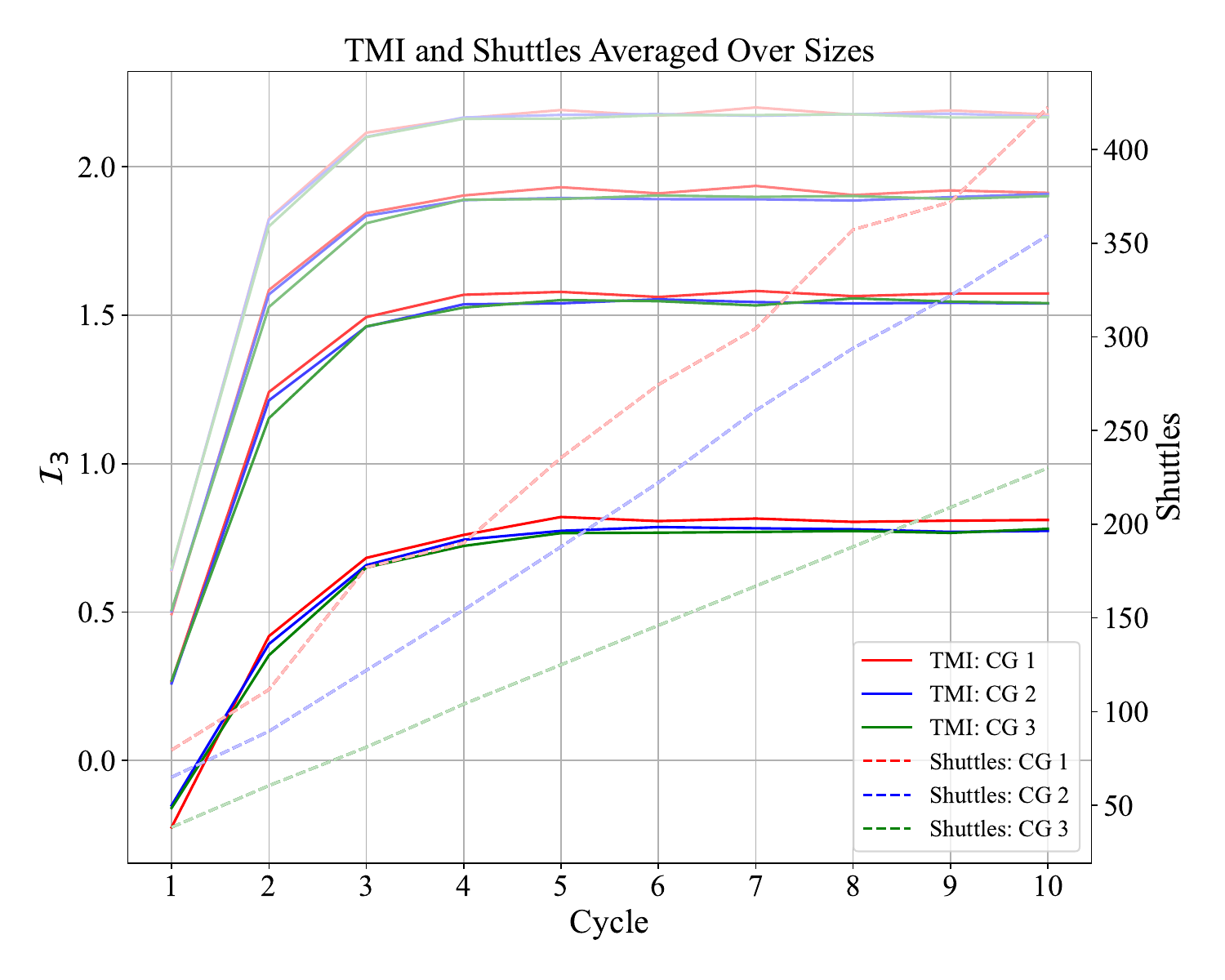}
\label{fig:crosstalkb}
\end{subfigure}
\caption{Results from the tensor network simulation of the $\llbracket 4,1,2 \rrbracket$ surface error-detecting code for a range of error rates, (a) with crosstalk (parameterized by $p_{1}^{r}$ after two-qubit gates) and (b) without crosstalk.}    
\label{fig:crosstalk}
\end{figure*}

As discussed in \cref{section:background}, \emph{crosstalk} refers to a problematic noise source that constrains the degree to which parallelization of noisy operations is possible in most current quantum devices \cite{murali2020softwarecrosstalk}. In many qubit technologies and their accompanying architectures, crosstalk can even arise from single-qubit operations \cite{sarovar2020detectingcrosstalk}. While these deleterious effects can be mitigated experimentally to some extent in single-qubit operations, crosstalk from two-qubit gates remains a significant challenge and is not well understood in general \cite{parrado2021crosstalk,sarovar2020detectingcrosstalk}. Given these issues, we aim to provide a preliminary investigation of crosstalk using a naive model; we simulated the impact of crosstalk originating from two-qubit interactions using the same tensor-network approach as utilized in \cref{fig:tensor_network_results}. 

The error model employed for the crosstalk simulations takes the following form, in addition to the noise parameters from $\mathcal{E}_{\text{TN}}$:

\begin{itemize}
\item After a two-qubit gate $CU$, we apply a CPHASE$(\zeta)$ gate with probability $p_{\text{cross}} = 10(p^{r}_{1} / 3\xi )$. Either the control or target qubit of the CPHASE$(\zeta)$ is designated from among the operands of the $CU$ gate, while the other qubit is randomly selected from the remaining, non-operational qubits.
\end{itemize}

One may inquire as to why we included the term $\xi$ in our modeling of the crosstalk. This term is used in order to take into account the connectivity differences on average across all CGs tested, as it is known that crosstalk is more prevalent in highly-connected devices \cite{parrado2021crosstalk,murali2020softwarecrosstalk}. Additionally, our crosstalk simulation adheres to one of the criteria on the definition of a crosstalk error \cite{sarovar2020detectingcrosstalk}; namely, our model violates the \emph{locality condition}, since one of the operands is scrambled to another random qubit with some probability. Using these model parameters, we analyzed the impact of crosstalk across a spectrum of gate errors, specifically using the probabilities $p^{r}_{1} = \{0.01, 0.03, 0.05, 0.08\}$, as discussed in \cref{section:spin_qubit_archs}. 

\begin{table*}
    \centering
    \begin{tabular}{|c|c|c|c|c|c|c|c|c|}
        \hline
        \textbf{CG} & \multicolumn{2}{c|}{\textbf{Error Rate 0.01}} & \multicolumn{2}{c|}{\textbf{Error Rate 0.03}} & \multicolumn{2}{c|}{\textbf{Error Rate 0.05}} & \multicolumn{2}{c|}{\textbf{Error Rate 0.08}} \\
        \cline{2-9}
        & Crosstalk & No Crosstalk & Crosstalk & No Crosstalk & Crosstalk & No Crosstalk & Crosstalk & No Crosstalk \\
        \hline
        1 & 0.6759 & 0.6501 & 1.4380 & 1.4000 & 1.7841 & 1.7337 & 2.0688 & 1.9844 \\
        2 & 0.6589 & 0.6308 & 1.4270 & 1.3728 & 1.7839 & 1.7161 & 2.0778 & 1.9777 \\
        3 & 0.6517 & 0.6192 & 1.4263 & 1.3684 & 1.7928 & 1.7118 & 2.0909 & 1.9725 \\
        \hline
        \textbf{Size} & \multicolumn{2}{c|}{\textbf{Error Rate 0.01}} & \multicolumn{2}{c|}{\textbf{Error Rate 0.03}} & \multicolumn{2}{c|}{\textbf{Error Rate 0.05}} & \multicolumn{2}{c|}{\textbf{Error Rate 0.08}} \\
        \cline{2-9}
        & Crosstalk & No Crosstalk & Crosstalk & No Crosstalk & Crosstalk & No Crosstalk & Crosstalk & No Crosstalk \\
        \hline
        4 & 0.6856 & 0.6616 & 1.4628 & 1.4123 & 1.8067 & 1.7439 & 2.0916 & 1.9984 \\
        5 & 0.7054 & 0.6720 & 1.4686 & 1.4327 & 1.8111 & 1.7591 & 2.0836 & 1.9999 \\
        6 & 0.6546 & 0.6316 & 1.4187 & 1.3743 & 1.7720 & 1.7180 & 2.0681 & 1.9735 \\
        7 & 0.6624 & 0.6364 & 1.4263 & 1.3828 & 1.7808 & 1.7225 & 2.0696 & 1.9759 \\
        \hline
    \end{tabular}
    \caption{The upper-half table shows the average $\mathcal{I}_{3}$ by coupling graph for various error rates, both with and without crosstalk; subsequently, the lower-half table displays the average $\mathcal{I}_{3}$ by size for various error rates, also with and without crosstalk.}
\label{table:crosstalk_CG_&_size_table}
\end{table*}

In \cref{fig:crosstalk}, the $\mathcal{I}_{3}$ results from tensor-network simulations based on the same compiled circuits are shown, conducted over the range of previously specified error rates for two scenarios: (a) without crosstalk and (b) with crosstalk. In each subfigure, the four groups of three colored lines (red, green, and blue) correspond to each error rate in the range, with the lowest rate at the bottom (opaque) and the highest error rate at the top (translucent and faded). For reference, the lower set of lines in \cref{fig:crosstalka} essentially mirrors the results displayed in \cref{fig:tensor_network_resultsa}. Notably, as we go from the lower error rates (at the bottom line group) in the range to the highest (top line group), the $\mathcal{I}_{3}$ growth rate progressively declines, indicating a non-linear relationship between error rates and $\mathcal{I}_{3}$. 

Since it could be difficult to clearly distinguish subtle, but nonetheless important, differences in \cref{fig:crosstalk}, we have additionally placed the average $\mathcal{I}_{3}$ received from all error rates (per CG and lattice size) in \cref{table:crosstalk_CG_&_size_table}. Taking stock of the data from \cref{table:crosstalk_CG_&_size_table,fig:crosstalk}, we can surmise that, as the error rates are increased with crosstalk, the resulting values calculated show a severely weakened advantage of CG$_{3}$ over the other two connectivities; this was not observed so prominently in our simulations without crosstalk. Indeed, at higher error rates, the $\mathcal{I}_{3}$ calculated for CG$_{3}$ even becomes worse than that of the other two CGs, indicating that the advantage of more local connectivity in CG$_{3}$ is counterbalanced by the introduction of more crosstalk.

%--------------------------------------------%
\section{Discussion} \label{section:discussion}

There are many points of discussion here. Firstly, let us consider the simulation for generating the AME(6,2) state and subsequent measurement of the Bell operator, in addition to the ESP. In particular, one may ask why we have chosen to graph $\langle B \rangle$, ESP, and the shuttle count together on one vertical axis, as we have done in \cref{fig:ame_results}; after all, the actual absolute values of the data we obtained lie staunchly on different orders of magnitude, and, a simple glance at the equations governing ESP and $\langle B \rangle$ seems to imply that no a priori relationship between these metrics exists; in this sense, it may appear that we have manufactured a correlation where in fact none exists. However, the parameters of the simulation were constructed in a way such that the only significant allowed change to every circuit per trial is the shuttle count. In addition, the major theme of our simulations involves the notion of constructing a \emph{hierarchy of comparison}, instead of establishing absolute values for our measures. In this way, our purpose centers on establishing relations to assess the quality of a spin-qubit architecture. As such, we graphed these three measures together in order to highlight a trend; namely, that the change in shuttle count throughout the simulation wrought changes that are observable in the Bell operator and ESP, and we observe exactly this behavior throughout our investigation. 

Next, one may critique the minor differences present in the ESP and Bell operator values reported in \cref{fig:ame_results}; in particular, one may state that the degree of change in $\langle B \rangle$ is not significant, and that furthermore, the ESP values themselves are very close to zero. In this work, we chose to analyze architectures on the basis of their entanglement properties; quantum entanglement structure is fragile in and of its own accord, and previous studies have shown that even small deviations in the Bell operator can signify large changes in qubit-qubit correlations, in addition to the general structure of multipartite entanglement \cite{quantumcirc_mme_states}. This effect also is present in the other metrics chosen, as very small differences in the $\mathcal{I}_{3}$ (or in the bipartite mutual information) can signal the advent of measurement-induced phase transitions \cite{tripartite1,tripartite2,bipartite1,bipartite2}; we will comment more on this shortly.

\begin{figure*}
\centering
\begin{subfigure}[b]{0.325\textwidth}
\centering
\includegraphics[width=\textwidth]{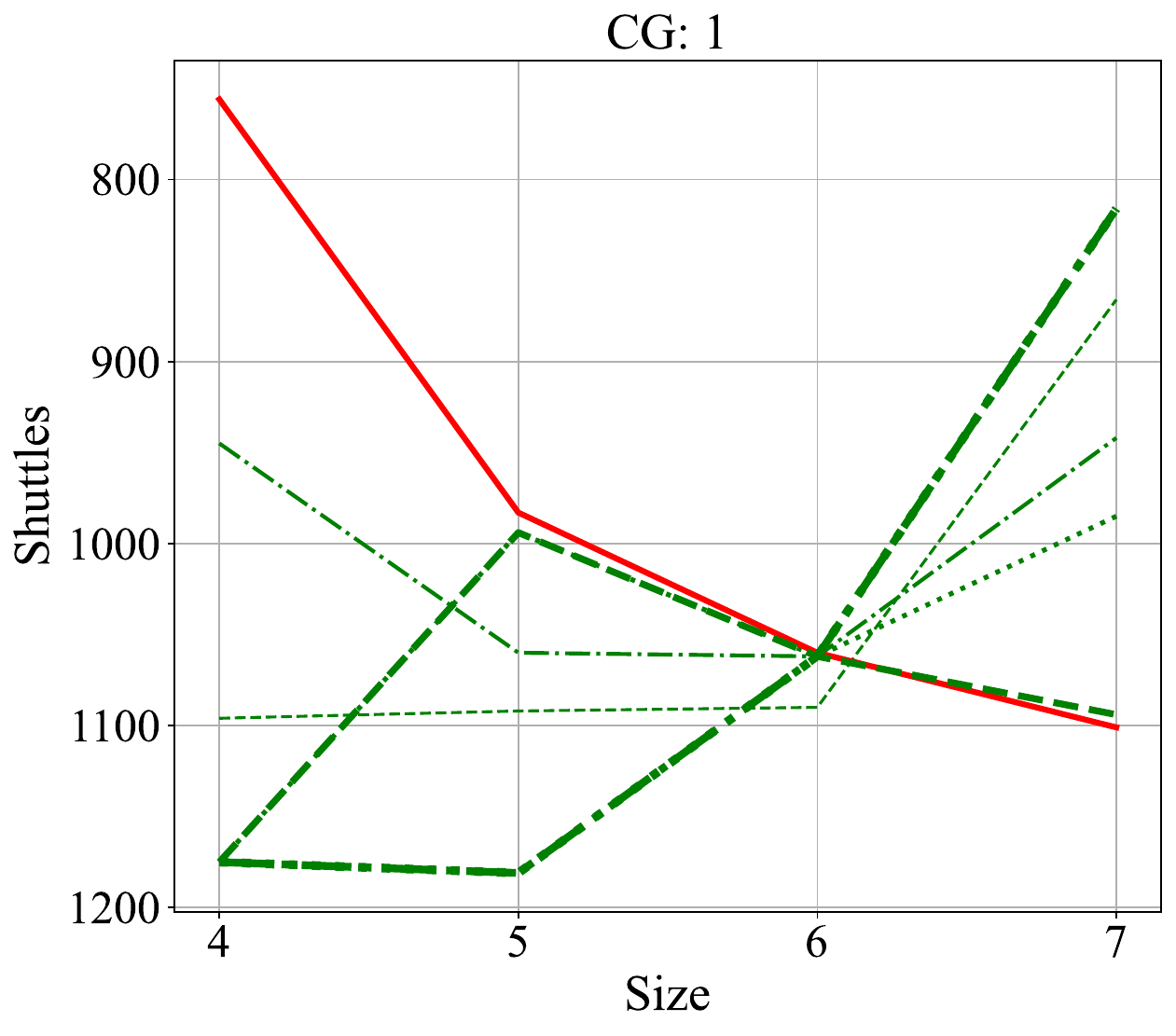}
\label{fig:AMEshuttlesA}
\end{subfigure}
\hfill
\begin{subfigure}[b]{0.325\textwidth}
\centering
\includegraphics[width=\textwidth]{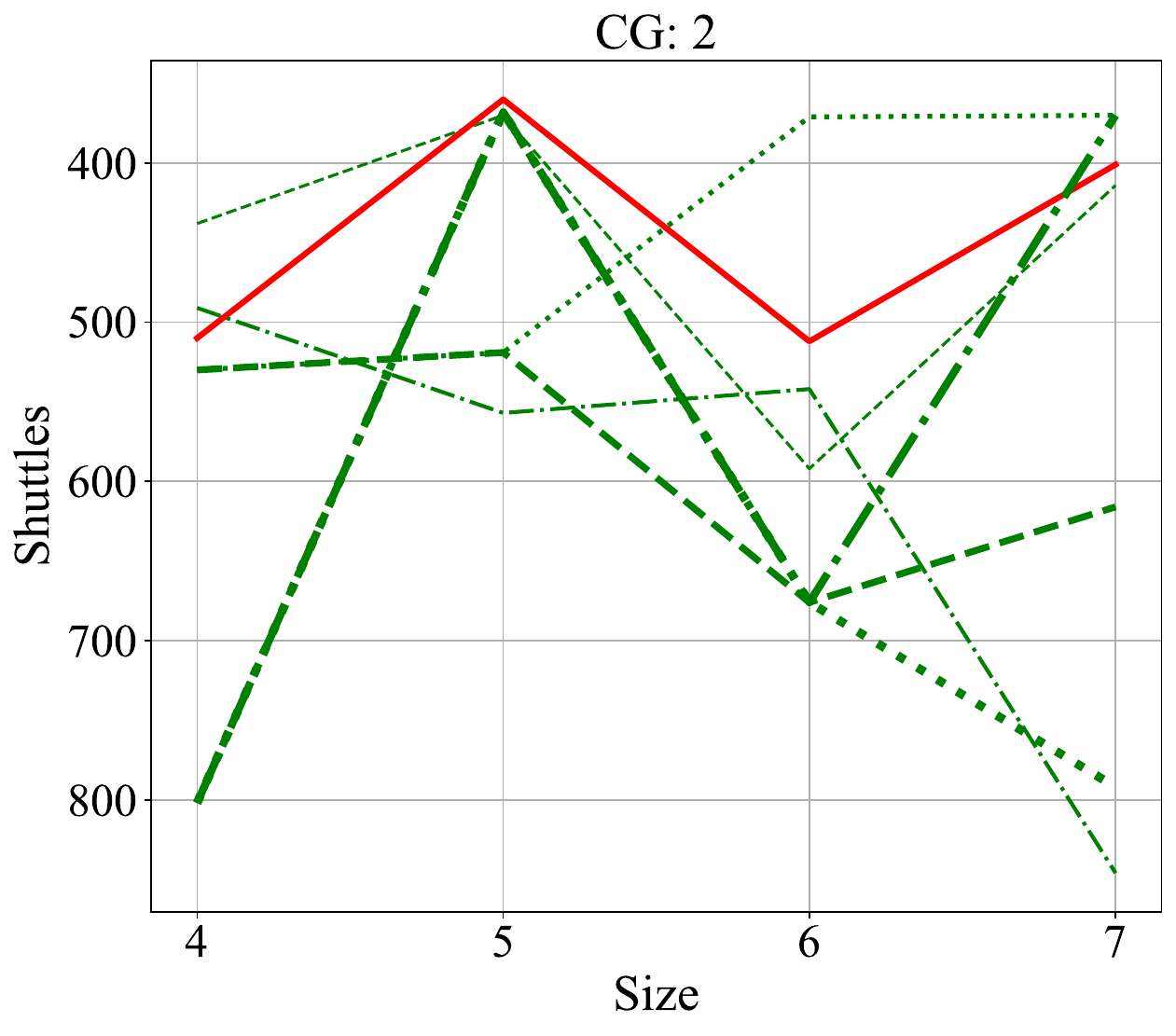}
\label{fig:AMEshuttlesB}
\end{subfigure}
\hfill
\begin{subfigure}[b]{0.325\textwidth}
\centering
\includegraphics[width=\textwidth]{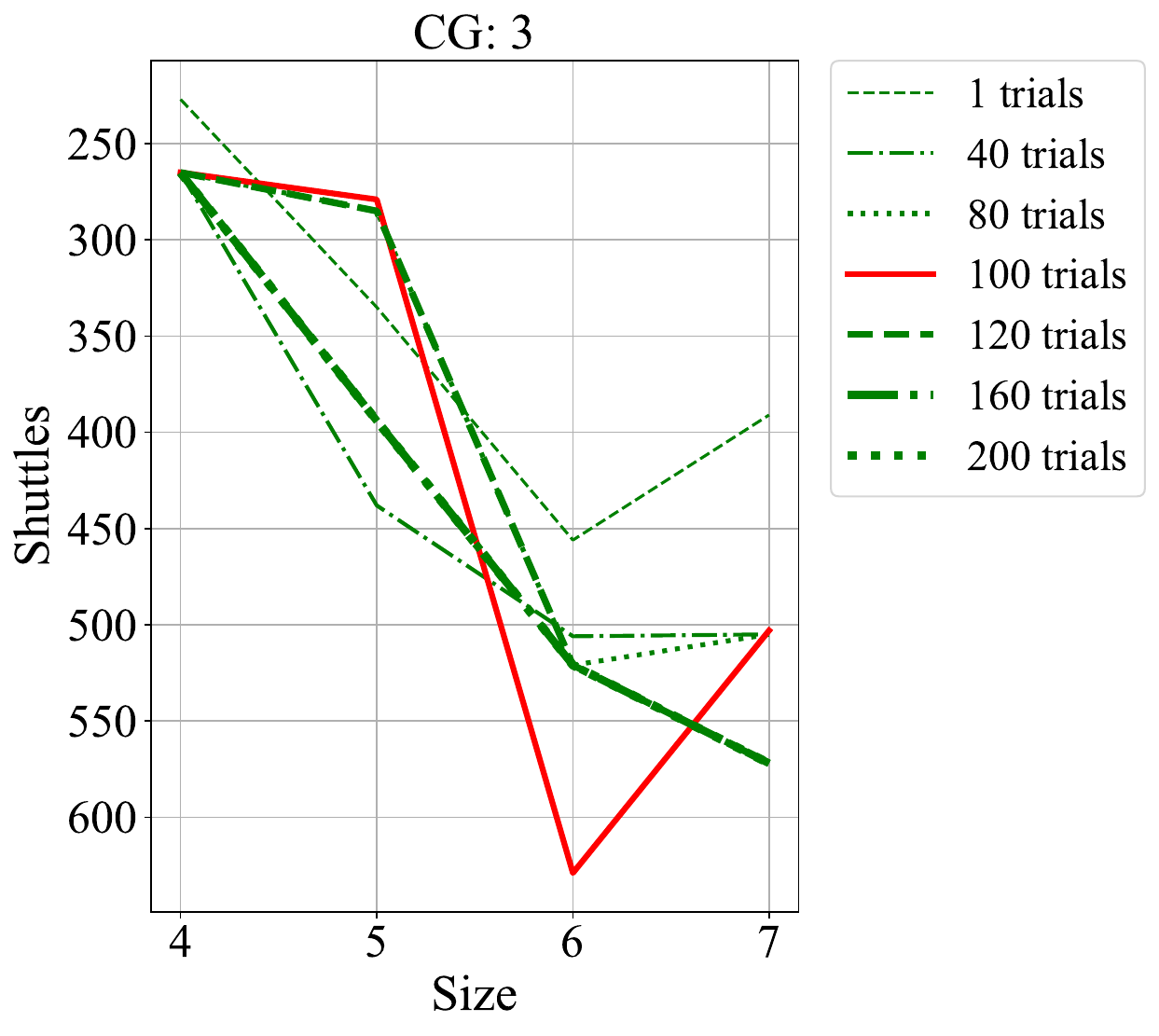}
\label{fig:AMEshuttlesC}
\end{subfigure}
\caption{Shuttle counts resulting from compiling the AME(6,2) circuit for each CG, using a range of initial placement optimization trials from SABRE, are shown. The red line represents the shuttle counts from simulations in \cref{fig:ame_generating_circuit}. Due to SABRE's limitations in optimizing shuttle operations, the resulting shuttles differ significantly. Typically, larger CG sizes allow initial placement algorithms to achieve more effective initial placements. However, in this case, we observe a potential trend in the opposite direction even with more trials.}
\label{fig:AMEshuttles}
\end{figure*}

In \cref{fig:ame_resultsb}, we observed incongruent behavior of the measures for CG$_{2}$; although all of the metrics are clearly correlated with one another, we discussed in \cref{section:ame_state_results} one possible reason as to why CG$_{2}$'s results fluctuate in comparison to those of the other architectures tested. In particular, this discrepancy can be explained by considering the initial placement algorithm SABRE. To better demonstrate the effect of initial placement with SABRE on the same circuit simulations of \cref{fig:ame_results}, a range of optimization trials (in between [1,200] trials) were tested, and the resulting shuttle counts were plotted in \cref{fig:AMEshuttles}. Here, we define a \emph{trial} in SABRE as a simultaneous three-step search process; more details on this can be found in \cite{li2019tackling}. As is evident, a higher or lower number of trials does not always equate with a more or less favorable shuttle count (even if the qubit density decreases with larger sizes), since SABRE is not optimized for shuttle operations present in spin-qubit devices. As explained in \cref{section:compilation_background}, the results can, therefore, vary significantly, regardless of how many trials are taken. Consequently, one can still conclude that compilation methods can have a significant influence on the values of all measures utilized in our studies; this agrees with the findings in literature \cite{kusyk2021survey,10313857}.

Regarding the simulation results shown in \cref{fig:logical_success_rates_results,fig:logical_qubit_ESP_results}, the close correspondence between both sets of data suggests that small quantum error-detection experiments with up to five cycles of stabilizer measurements could be possible using any of the devices considered in this work. CG$_{3}$ outright obtains the highest value for both the ESP and logical success rate in this cycle range, with CG$_{2}$ and CG$_{1}$ falling in line with lower logical success rate and ESP values. By examining \cref{fig:logical_qubit_ESP_resultsa} and \cref{fig:logical_qubit_ESP_resultsb} more closely, we observe the same partial ordering as per the values of both of the measures captured, with CG$_{3}$ showing an advantage over the other two as expected. However, for certain sizes, both the metrics of CG$_{2}$ and CG$_{3}$ converge closely. The main message from this observation is that, when averaged over many cycles, the measured ESP and logical success rates for CG$_{2}$ and CG$_{3}$ do not differ heavily; this is significant, as it implies (as in \cref{fig:ame_results}) that for near-term experiments with highly-entangled quantum states, it is not necessarily more beneficial to fabricate a device with higher local connectivity. Instead, under the assumption of utilizing quantum compilation methods, we note that realizing a properly-compiled quantum circuit on CG$_{2}$ in fact approaches the values of CG$_{3}$ in several key metrics that we have studied. We also observe this for \cref{fig:tensor_network_results,fig:crosstalk}, although as the crosstalk effect is increased, we notice that the $\mathcal{I}_{3}$ calculated for CG$_{2}$ and CG$_{3}$ become even worse than CG$_{1}$. 

Looking more deeply at the ESP and logical success rate, it becomes clear that the ESP approaches 0 relatively quickly while $p_s$ converges to approximately 0.25. As explained in \cref{section:esp_background}, ESP stands out from the other metrics due to its worst-case approach to estimating performance and with better computational efficiency. In contrast, the logical success rate $p_s$ is calculated over numerous state-vector simulation trials; one can expect that, under the conditions affecting our noise model, projective measurements should repeatedly project the logical state back into the codespace, as was discussed in  \cref{section:logical_success_results}. Although both metrics scale differently with larger circuits, their relative performance remains correlated, as previously discussed, making them equally reliable for their respective use cases.

We also remark that the noise models utilized for the tensor-network simulations of \cref{fig:tensor_network_results,fig:crosstalk} indicate a distinct noise model from those of \cref{section:ame_state_results,section:logical_success_results}. Although the inclusion of idling errors do not affect the main conclusions from our results, the crosstalk present in our other error model does affect the conclusion of our study. This effect can be seen in \cref{fig:crosstalka}; as the error rate is incremented, the $\mathcal{I}_{3}$ grows non-linearly, per cycle. At the higher error rates (shown in faded colors), we discern that the three size-averaged CGs converge after smaller and smaller cycle numbers; this indicates the effects of crosstalk at higher error probabilities, since all other noise effects in our simulations are accounted for. Taken at face value, these results further imply fundamental limitations to the size and depth of quantum circuits that the selected architectures will be able to perform. 

Regarding the results shown in \cref{fig:tensor_network_results}, we mentioned the possible appearance of a \emph{measurement-induced phase transition} (MIPT) from volume- to area-law phases. While it is known that the $\mathcal{I}_{3}$ generally is a good indicator of such MIPTs, we assert that more investigation is needed in order to ascertain this supposition; for a start, the size of the circuits tested (in this case, the surface code circuit) could be scaled with the size of the device, in order to maintain roughly equal qubit densities. Additionally, both the entanglement entropy and the $\mathcal{I}_{3}$ should exhibit the prototypical peaks consistent with MIPTs, and our results do not confirm this trend at the moment, even though other works have observed the existence of this phenomenon in the surface code \cite{scs_entanglement_phases}. We leave the exploration and confirmation of MIPTs in such prospective architectural studies for future work. 

The results from \cref{fig:crosstalk} and \cref{table:crosstalk_CG_&_size_table} support our findings that, in the presence of crosstalk, there are underlying tradeoffs with regard to adding local connectivity to a spin-qubit device. In particular, it is evident that, even at the error rate $0.03$, the size-averaged $\mathcal{I}_{3}$ are the same for both CG$_{2}$ and CG$_{3}$. As we increase the error rate for the crosstalk simulations, CG$_{3}$ attains the worst $\mathcal{I}_{3}$ of the three connectivity graphs. These same results are visible, even without crosstalk, but become accentuated as we add crosstalk to error model $\mathcal{E}_{\text{TN}}$. As such, we can deduce that, in the presence of the naive crosstalk model used here, our results indicate a disadvantage when utilizing the most highly-connected CGs. It is interesting to remark that we have also probed the same simulations, but with single-qubit crosstalk parameters added; again, the same tendency is again present, albeit to larger degree. 

Surprisingly, the CG-averaged size results from \cref{table:crosstalk_CG_&_size_table} imply that lattice size $2 \times 6$ on average achieves consistently the lowest $\mathcal{I}_{3}$. One may ordinarily suspect that the largest device, the $2 \times 7$ array, would achieve the most favorable $\mathcal{I}_{3}$, as all qubits in the circuit are relatively isolated from one another; however, we do not observe this in our results. One intuitive reason may have to do with the qubit density of the array. Consider, for example, an empty spin-qubit array; as we start to add qubits, the degree to which they can be correlated with one another is quite limited, as it is known that entanglement correlations exponentially decay over small distances when exposed to environmental effects \cite{yu2009suddendeath}. Notwithstanding, we will eventually approach a \emph{critical point} of filling the lattice, after which it becomes progressively more and more difficult to perform shuttling operations, and the benefit of doing so in a spin-qubit device fades. This phenomenon is known in percolation theory as the \emph{percolation threshold}, and separates classical ``phases" of particle behavior into those which interact independently versus those which are strongly dependent upon the interactions of neighboring particles or clusters on the lattice \cite{percolation_two_dim}. For a Euclidean square lattice, the site-percolation threshold is known to be $p_{\text{perc}} \approx 0.5927$, for which there is an approximately $1\%$ difference with the qubit filling in our $2 \times 6$ array (i.e., $7/12 \approx 0.583$) \cite{percolation_two_dim}. In light of this, it seems reasonable to conclude that, for the seven-qubit circuits that we have tested, a $2 \times 6$ bilinear array may yield better $\mathcal{I}_{3}$ results. One may be able to formalize this concept more, by looking concretely into \emph{critical exponents} and how they affect the quality of entanglement on the device as it realizes the quantum circuit; more investigation is needed into this, which we leave for future work. In the most realistic case, a pragmatic experimentalist would consider a $2 \times 6$ array as a subset of a larger spin-qubit array, permitting the system to be always operated at an optimal filling. This condition allows for the exploration of optimally-sized lattices for circuit execution in future spin-qubit architectures. 

We would like to briefly note that our simulations, at first glance, do not seem scalable (with the notable exception of ESP, which scales as $\mathcal{O}(g)$, where $g$ is the number of gates in a given circuit). One reason for this issue may be due to necessity of large-scale Monte Carlo simulations, tensor-network contraction, or calculations such as matrix diagonalization, which are known to scale at worst $\mathcal{O}(n^{3})$ for dense matrices (where $n$ is the dimension of the matrix), and at best $\mathcal{O}(n)$ for sparse matrices. However, our methods could be substantially improved by utilizing optimized tensor-network contraction techniques, such as those from \cite{gray2021hyper,gray2024hyperoptimized}. We leave such optimizations for future work.

Finally, one may be tempted to establish a global precedent of lattice sizes, given the overall implications of our work; after all, only one set of results from our work (\cref{fig:ame_resultsb}) does not immediately agree with the conclusions drawn from the rest. Accepted as is, we cannot claim that a particular lattice size of spin-qubit device would be more or less advantageous than another. As stated before, the major reason for this added nuance to our results lies in the fact that the initial placement algorithm, SABRE, is not optimized for all of the architectural features that typically characterize spin-qubit technologies. If an initial placement algorithm for spin-qubit architectures can be developed with provable guarantees for solution quality (such as a specialized version of \cite{variational_compiler_qec} for spin-qubit technology), we believe that the results of our work should be revisited. However, in spite of the complications discussed above, we can safely conclude that circuit compilation does greatly influence the four metrics that we have proposed, in order to study the connectivity tradeoffs between future architectural designs. In light of our demonstration, it is highly probable that, by utilizing appropriate quantum circuit-compiling techniques, it is possible to achieve better metric values for more sparsely-connected devices. In particular, our results strongly suggest that, while CG$_{2}$ and CG$_{3}$ indeed may not have such different entanglement properties, the extra connectivity of CG$_{2}$ and CG$_{3}$ do not appear to be worth the fabrication effort, especially in light of the convincing results evinced by our crosstalk simulations. These outright worse results were prominent in almost all of the data obtained, especially in the limit of both higher cycle numbers of stabilizer measurements, as well as higher lattice sizes (i.e., lower qubit densities) when we employed our crosstalk model. 

%--------------------------------------------%
\section{Conclusion} \label{section:conclusion}

In this work, we have presented a framework based on quantum information-theoretic and compilation-based measures for methodically evaluating the entanglement properties of prospective quantum architectures. More specifically, we have utilized the following entanglement measures: the average \emph{Bell operator} $\langle B \rangle$; the \emph{logical success rate $p_{s}$} for the smallest error-detecting surface code; and the \emph{tripartite mutual information} $\mathcal{I}_{3}$ (again for the $\llbracket 4,1,2 \rrbracket$ surface code). We also proposed a modification to the \emph{estimated success probability}, a known metric in quantum compilation, in order to take into account the effects of decoherence in certain quantum devices stemming from \emph{thermal relaxation}. We demonstrated the usefulness of our techniques by realizing an architectural study which profiles the structure of entanglement of generated quantum many-body states. Indeed, it is supposed that more local connectivity is immediately beneficial towards executing highly-entangled quantum circuits. Surprisingly, we find that, under the assumptions of the noise models chosen, it is possible to approach comparable qubit-qubit correlations, ESP, logical success rates $p_{s}$, and $\mathcal{I}_{3}$ values to the most highly-connected quantum devices in our set, by utilizing appropriate techniques from spin-qubit quantum compilation with SpinQ \cite{SpinQ} and beSnake \cite{beSnake}, as well as the advantage conferred through the usage of shuttling. Our results suggest that for small-scale spin-qubit experiments, more device connectivity does not necessarily guarantee an improvement in the quality of quantum entanglement arising from circuit-prepared quantum states; this narrative was strongly apparent in our $\mathcal{E}_{\text{TN}}$ error model simulations, as well as for all other simulations (albeit to a somewhat lessened degree, which we attribute mainly to the lack of a specialized initial placement algorithm). Central to our approach was the incorporation of expected device characteristics, such as gate and shuttling durations and expected dominant noise sources. 

Our results come in the wake of several recent works attempting to evaluate the device-level fitness of spin-qubit architectures for near-term quantum error-correction experiments \cite{gozde_article_floquet_qec,siegel2024towards,helsen2018quantum}. Modification of our methodology for the direct simulation of such error-correction schemes is straightforward, and in this way, our framework can assist with device prototyping and design, without the financial, manufacturing, and time overhead costs normally ascribed to the development of new experimental quantum devices. Indeed, the framework proposed can be utilized to test further spin-qubit architectures, outside of the connectivities tested in this work. Moreover, by modifying parameters of the simulation such as gate duration or the parameters of the compiler, one may be able to investigate other qubit technologies as well, such as trapped-ion, neutral-atom and superconducting devices. One may be able to simulate more nuanced environmental hardware noise sources using a \emph{density-matrix} simulation \cite{dicarlo_sc_dmsim}; we leave this for future exploration.

As we remarked in \cref{section:discussion}, several future directions are possible. We address a few of these ideas below: 

\begin{itemize}
\item As we have investigated a simple quantum error-detection code's logical success rate, it would also be useful to simulate directly the logical success rate of various transversal logical operations in the $\llbracket 4,1,2 \rrbracket$ surface code, as was experimentally realized in \cite{experiment2} for superconducting qubit technologies. This concept would help to inform what logical-level success rates can be expected for similar experiments in near-term spin-qubit devices.
\item The development of a specialized spin-qubit initial placement algorithm would help to substantiate and bolster many of the conclusions that we drew from our study, as we have mentioned in \cref{section:discussion}.
\item \cite{linke2017experimental} attempted to address the question of connectivity among different types of qubit technologies, focusing on the implementation of various quantum algorithms in trapped-ion and superconducting devices available at the time. As the semiconductor spin-qubit community scales up their devices, it may be useful to revisit this work and benchmark again, with near-fault-tolerant logical-level algorithms and specialized compilation techniques (such as those used in our work).
\item Larger circuits could be leveraged in order to investigate the presence of an MIPT, which we preliminarily have observed in our results with tripartite mutual information. By scaling the size of a commensurate error correction code, one may be able to investigate the asymptotic threshold properties that emerge, as in \cite{helsen2018quantum} (although such efforts may be problematic, as the definition of a threshold in QEC is contingent on the existence of arrays in the asymptotic limit). Scaling the size of our simulations would be expensive computationally; however, there are many methods by which one can optimize tensor-network simulations for large-scale quantum circuit investigations \cite{gray2021hyper,gray2024hyperoptimized}.
\item It has been suggested in \cite{undseth1_multi} that multiple device cores could be connected via shuttling- and microwave-based module interlinks \cite{baart2016single, dijkema2023two}; it would be fascinating to see whether our observations hold in this modular regime.
\item We have neglected gate parallelization in this work, in order to focus on the unique characteristics of spin-qubit devices: namely, finite-connectivity architectures and qubit-density, combined with allowed shuttling operations. In the future, one could benefit from investigating parallelization protocols in this vein.
\item Lastly, one could easily expand this study towards even more practical techniques related to near-fault-tolerant experiments, such as other types of quantum error correction codes \cite{fan2024overcoming,old2024lift,roffe1,siegel2024towards} or \emph{magic-state distillation} \cite{souza2011experimental,rodriguez2024experimental,msd0}, ingredients that are needed for large-scale, fault-tolerant quantum computing.
\end{itemize} 

The architectural conclusions that we have arrived at in this study are specifically bound to the parameters that have been chosen for each error model. Although we strongly suspect that our observations hold in more general circumstances, future work will be needed in order to ascertain the specificity of our results for spin-qubit platforms. As an example, the ratio of the shuttling time to the coherence time may have a significant impact on the inferences arrived at. 

%--------------------------------------------%
\section{Acknowledgments}

%--------------------------------------------%
\subsection{People}

We thank Alexander Ivlev and G\"{o}zde \"{U}st\"{u}n for useful comments.  

%--------------------------------------------%
\subsection{Funding}

MS and SF thank the Intel Corporation for financial support. AS acknowledges funding from the Dutch Research Council (NWO) through the project ``QuTech Part III Application-based research" (project no. 601.QT.001 Part III-C—NISQ). B.U. acknowledges support from the “Quantum Inspire–the Dutch Quantum Computer in the Cloud” project (Project No. NWA.1292.19.194) of the NWA research program
“Research on Routes by Consortia (ORC),” which is
funded by the Dutch Research Council (NWO). X.X. acknowledges support from the NWO via the National Growth Fund programme Quantum Delta NL (Grant No. NGF.1623.23.024).

%--------------------------------------------%
\subsection{Software Resources}

The following libraries were utilized in this project: SpinQ \cite{SpinQ}; beSnake \cite{beSnake}; Qiskit \cite{qiskit2024}; and quimb \cite{gray2018quimb}. The compilation of all circuits and the state-vector simulations were executed on an Intel® Core™ i9 processor 14900HX with 32GB of RAM, and the tensor network simulations were run on a multi-node cluster server.

%--------------------------------------------%
\subsection{Author Contributions}

NP, MS, and AS developed the quantum circuits and quantum-information-theoretic measures used to benchmark the architectures, with input regarding near-term spin-qubit experiments from BU and XX. NP, AS, and MS conceptualized the noise models, with assistance from LV, BU, and XX concerning realistic hardware considerations. NP implemented the code for the compiler and architectures, and executed all tensor-network simulations with the help of MS. AS, LV, and SF coordinated the project goals, supervised the project, and provided useful insights during the writing process.

%--------------------------------------------%
\clearpage
\bibliography{bibliography}

\clearpage
\appendix 

%--------------------------------------------%
\section{Success rates for selected simulations with Pauli error rates} \label{section:appendix_logical_success_with_pauli_errors}

\begin{figure*}
\centering
\begin{subfigure}[b]{0.325\textwidth}
\centering
\caption{ }
\includegraphics[width=\textwidth]{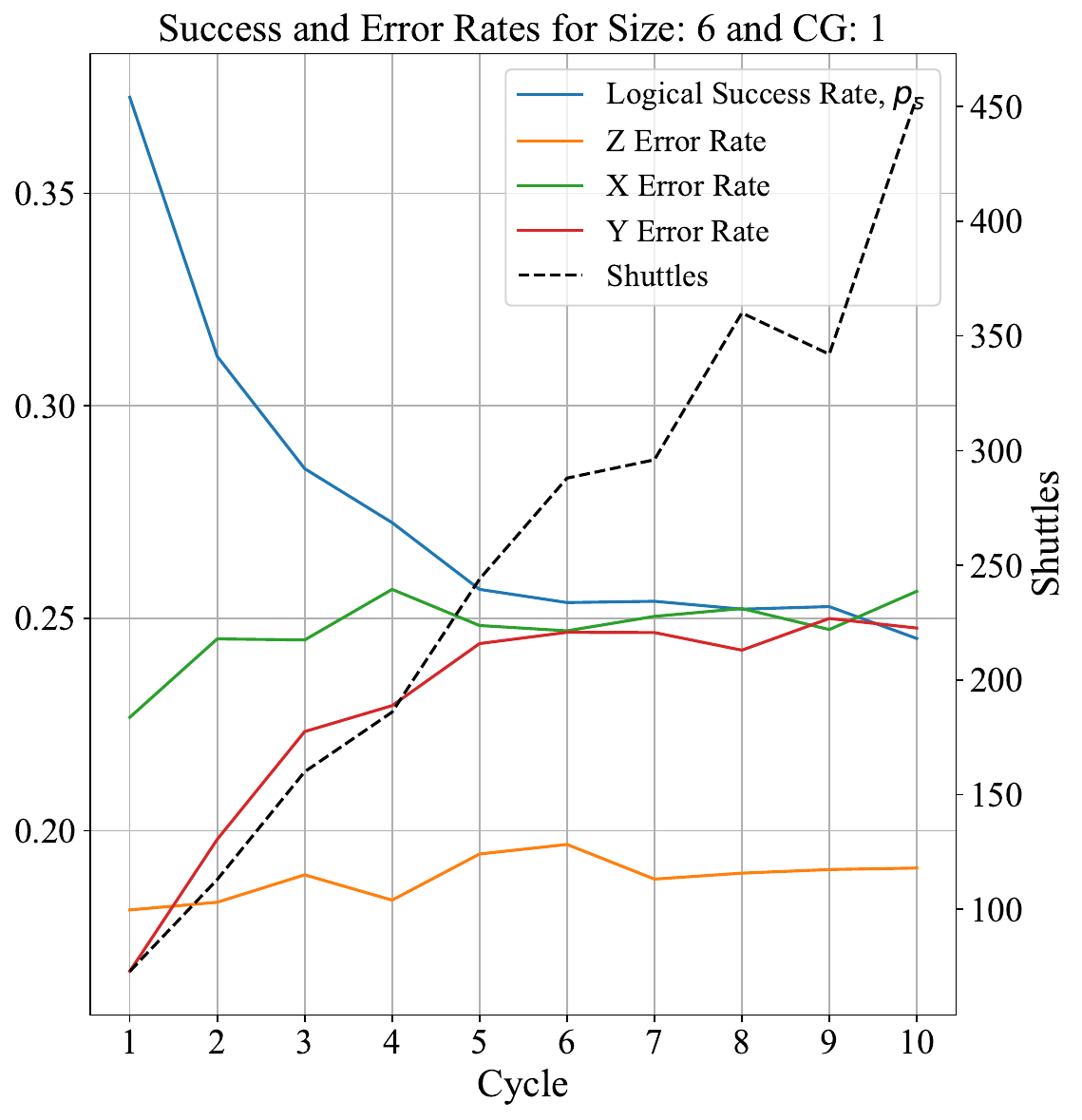}
\label{fig:logical_examplesa}
\end{subfigure}
\hfill
\begin{subfigure}[b]{0.325\textwidth}
\centering
\caption{ }
\includegraphics[width=\textwidth]{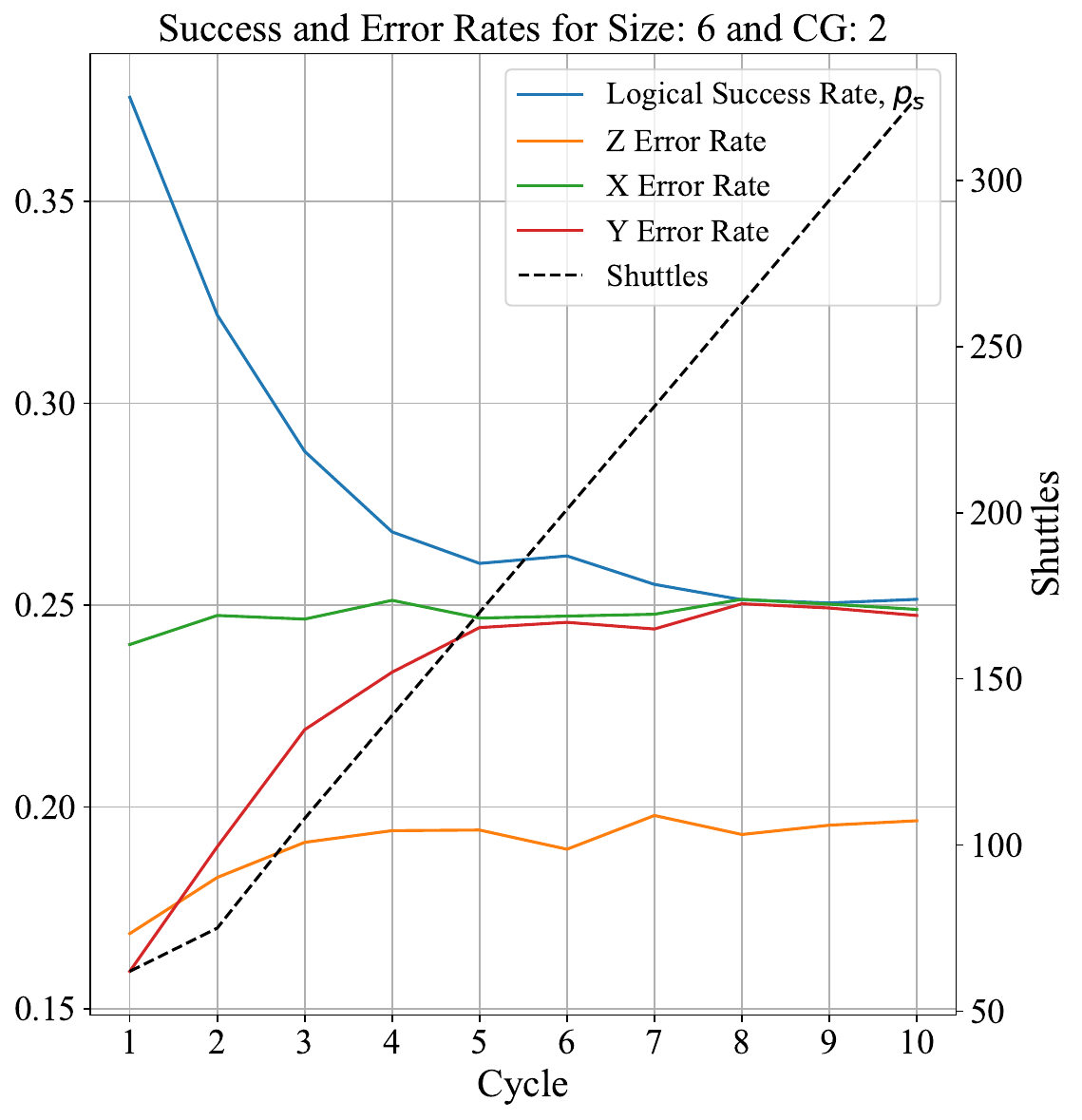}
\label{fig:logical_examplesb}
\end{subfigure}
\hfill
\begin{subfigure}[b]{0.325\textwidth}
\centering
\caption{ }
\includegraphics[width=\textwidth]{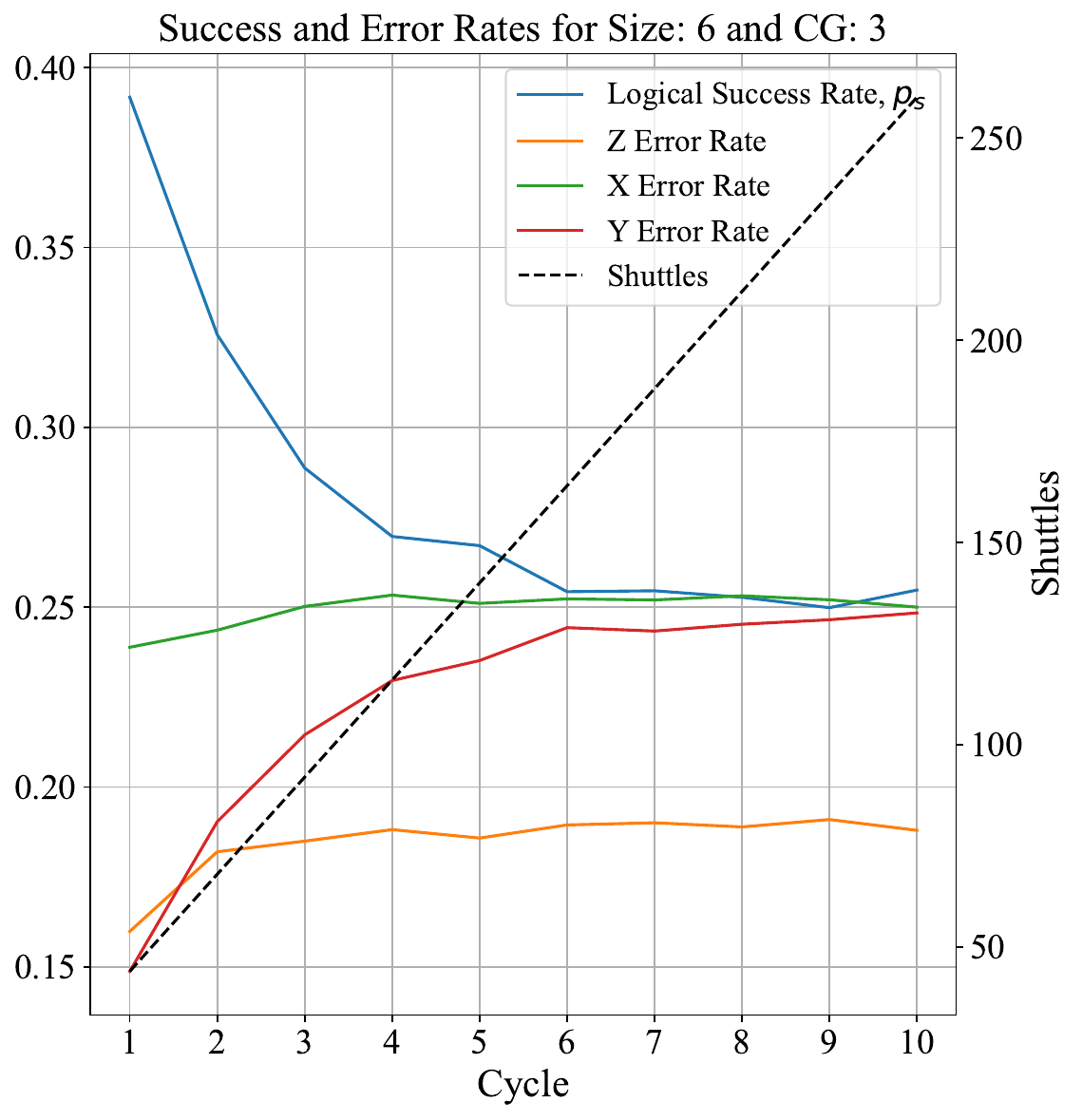}
\label{fig:logical_examplesc}
\end{subfigure}
\caption{Success rates for select simulations with Pauli error rates. (a) Architecture with CG$_{1}$ of size $2 \times 6$ (b) Architecture with CG$_{2}$ of size $2 \times 6$ (c) Architecture with CG$_{3}$ of size $2 \times 6$.}
\label{fig:logical_examples}
\end{figure*}

In the results shown in \cref{fig:logical_qubit_ESP_results,fig:logical_success_rates_results}, the success rate is calculated as the frequency of $000$ or $111$ syndrome measurement results, relative to the total number of Monte Carlo simulation trials. In \cref{fig:logical_examples}, we present individual success rates alongside the corresponding Pauli error rates for each of the three CGs of size 6. The differences observed in the figures are subtle; however, with increased connectivity, we see a reduction in shuttle operations and a modest increase in the overall success rate.

Examining the Pauli error rates more closely, we noticed that Z and Y errors increase slightly up to approximately cycle 4, after which they stabilize. Conversely, Y errors exhibit a more pronounced increase until reaching a plateau at the same level as X errors. Thus, if Z and X error rates are relatively stable, Y errors appear to be the primary factor impacting the logical success rate. This is evident from the distinct trend between Y errors and the success rate. These findings suggest that the combination of this circuit with the specific architectures and error model is particularly susceptible to Y errors, and with our framework, we are able to explicitly identify and extract this vulnerability, providing valuable insights into the system's behavior.

From a more detailed perspective, Z errors are relatively low because they occur exclusively with a \(100\) stabilizer measurement. Additionally, the error model is negatively biased towards Z gates due to the effects of decoherence: Z errors introduced from the decoherence-induced errors can be canceled by subsequent Z errors occurring from the depolarizing model. In contrast, X and Y errors arise under different conditions: X errors occur when the stabilizer measurement is \(010\) or \(001\), while Y errors manifest in measurements of \(110\) or \(101\) which explains their higher error rate compared to Z errors. For the X errors, we see that, at the beginning of the simulation, the amount of relative X errors is similar to the proportion dictated by depolarizing noise; however, as the simulation progresses, the injection of other Pauli errors can cause non-trivial error mixing, thus resulting in slightly larger error rates of X errors than may be normally expected in the depolarizing noise framework.

\onecolumngrid

\end{document}